\documentclass[a4paper,fleqn,usenatbib]{mnras}

\usepackage[T1]{fontenc}
\usepackage{ae,aecompl}

\usepackage{graphicx}	
\usepackage{amsmath}	
\usepackage{amssymb}	

\title[E \& B modes in a field with multiple lenses]{Deep lensing with a twist: E and B modes in a field with multiple lenses}

\author[A. K. Bradshaw et al.]{
Andrew K. Bradshaw,$^{1}$\thanks{E-mail: akbradshaw@ucdavis.edu (AKB)}
M. James Jee,$^{2}$
J. Anthony Tyson$^{1}$
\\
$^{1}$Physics Department, University of California, Davis, 1 Shields Ave., Davis 95616, USA\\
$^{2}$Department of Astronomy, Yonsei University, 50 Yonsei-ro, Seoul 03722, Korea
}

\date{Accepted 2018-10-13. Received in original form 2017-09-27}

\pubyear{2018}

\begin{document}
\label{firstpage}
\pagerange{\pageref{firstpage}--\pageref{lastpage}}
\maketitle

\begin{abstract}
We explore the weak lensing E- and B-mode shear signals of a field of galaxy clusters using both large scale structure $N$-body simulations and multi-color Suprime-cam \& {\it Hubble Space Telescope} observations. Using the ray-traced and observed shears along with photometric redshift catalogs, we generate mass maps of the foreground overdensities by optimally filtering the tangential shear that they induce on background galaxies. We then develop and test a method to approximate the foreground structure as a superposition of NFW-like halos by locating these overdensities and determining their mass and redshift, thereby modeling the background correlated shear field as a sum of lensings induced by the foreground clusters. We demonstrate that the B-mode maps and shear correlation functions, which are generated by similarly filtering the cross shear in this method, are in agreement with observations and are related to the estimated cluster masses and locations as well as the distribution of background sources. Using the foreground mass model, we identify several sources of weak lensing B-modes including leakage and edge effects, source clustering, and multiple lensing which can be observed in deep cosmic shear surveys.
\end{abstract}


\section{Introduction}
Weak gravitational lensing provides a powerful tool to map and study the formation of large scale structure in the Universe. Foreground mass overdensities like clusters of galaxies cause light from distant sources to be deflected, causing extended objects to become tangentially sheared. Conversely, under-dense regions of space cause images of distant galaxies to appear radially aligned to the center of the void. Using this knowledge of the relationship between weak lensing shear and the gravitational potential, finite field non-linear reconstruction techniques can be applied to the shear to generate the locations and depths of potentials on the sky (\citet{Tyson1990ApJ}, \citet{Bartelmann1995A&A}, \citet{Kilbinger2015RPPh}).

By analogy with electromagnetism, \citet{Kaiser1992ApJ} and \citet{Stebbins1996astro.ph} showed that the shear field can be decomposed into two components: the E-modes and the B-modes (so-called E/B, gradient/curl, non-vortical/vortical, and scalar/pseudo-scalar perturbations). \citet{Hilbert2009A&A} and others have shown that to a good approximation lensing produces only E-modes when large samples of lenses are used, with lensing-induced B-modes having a much reduced amplitude which have been observationally consistent with zero. Accordingly, E-mode patterns have been the main focus of gravitational lensing studies due to their usefulness in measuring density perturbations. Unfortunately this has led to the treatment of B-modes as a contaminant, and their exclusive use as a probe of systematic error. However, there are numerous reasons to investigate B-modes as a source of astrophysical signal.

B-modes can be observationally generated through random or intrinsic alignments of source galaxies as well as through their inhomogeneous (clustered) distribution \citep{Schneider2002A&A}. Additionally, \citet{Kilbinger2006A&A} show that E- and B-modes can mix when there is a lack of close projected pairs of galaxies, and conversely that B-modes are induced by finite survey size. Finally, if two (or more) lensing clusters are projected along the line of sight, the lensing of a background E-mode produces a so-called double-lensing B-mode pattern as shown in maps of \citet{Bertin2001ApJ} and in the shear correlation functions of \citet{Cooray2002ApJ}. Every cosmological survey has some of these issues, implying their presence in all weak lensing maps and correlations. The amplitude and scale of these observationally-induced B-modes is determined by the depth and width of field and the distribution of mass within it. As we will demonstrate, modeling of the lensing field can yield useful signal from these inevitable B-modes.

Naturally, other observational issues related to the point spread function (PSF), such as atmospheric turbulence or optical distortions, can also induce spurious B-modes. For instance, it was shown in \cite{Guzik2005PhRvD} that calibration errors in seeing, depth or extinction in excess of 3\% r.m.s. would generate B-modes that bias the shear correlation function beyond statistical errors. For this reason, every weak lensing analysis addresses the issue of PSF modeling by using stars as a reference for the spurious PSF shear. Additionally, shear measurement methods must be calibrated to provide unbiased estimates under realistic conditions.

Understanding of the above sources of B-mode is critical if novel astrophysical sources of B-modes are to be sought. For instance, the intrinsic alignments of source galaxies induced by tidal gravitational interactions with their environment are an active area of research \citep{Kirk2015SSRv}, and weak lensing B-modes produced by these interactions present an observational opportunity to constrain models of galaxy formation and evolution \citep{Joachimi2013MNRAS}. Other more exotic cosmological models can generate B-modes through tensor-vector perturbations \citep{Stebbins1996astro.ph}. Alternative models can even break the fundamental assumption of statistical homogeneity and isotropy of sources, or alter the propagation of photons through anisotropic cosmologies (\citet{Kristian1966ApJ}, \citet{Pitrou2015PhRvD}) which have yet to be observationally excluded (\citet{Kristian1967ApJ}, \citet{Valdes1983ApJ}). Given the abundance of possible astrophysical sources of B-modes, it is worthwhile to conduct a comprehensive search for B-modes in cosmic shear surveys as a potentially useful signal instead of purely an indicator of systematic error.

Precision mass mapping and measurement of cosmic shear primarily necessitates a high number density of background galaxies to statistically reduce the shot noise from intrinsic galaxy ellipticity. For this reason, deep surveys of the sky have been specially designed to maximize the observed number density of background galaxies, decreasing the statistical uncertainty in the measured shear field by the square root of the observed number density of objects. Technology has accelerated the progress of weak lensing from first measurement in deep CCD images \citep{Tyson1990ApJ} to the discovery \citep{Wittman2001ApJ} and characterization of clusters (\citet{vonderLinden2014MNRAS}, LoCuSS \citet{Okabe2010ApJ}, CCCP \citet{Hoekstra2015MNRAS}, and CLASH \citet{Umetsu2016ApJ}) of and voids \citep{Melchior2014MNRAS} in galaxy surveys. Weak lensing can also be used as a tool for cosmology, either through a census of peaks in mass maps (\citet{Schirmer2007A&A}, \citet{Shan2018MNRAS}, \citet{Kacprzak2016MNRAS}, or through spatial correlations of the shear (as first measured by \citet{Bacon2000MNRAS}, \citet{Kaiser2000astro.ph}, \citet{vanWaerbeke2000A&A}, \citet{Wittman2000Nature} and more recently by \citet{Zuntz2017arXiv} \citet{Fu2014MNRAS} \citet{Jee2016ApJ} and \citet{Hildebrandt2017MNRAS}). The power of each of these weak lensing surveys depends on the size of the cosmological volume it can probe, which can be aided by going wider or deeper. One future survey, the LSST, will have a $5\sigma$ depth of $r\sim27$ (effectively hours of observing time on a 6.7 meter equivalent) across $18,000$ square degrees at the end of its nominal 10 year survey \citep{Zhan2018RPPh}, achieved by efficiently stacking hundreds of exposures of billions of galaxies in its 9.6 sq. degree field of view. 

In this paper we use observations of a small region of the sky imaged to LSST 10-year depth, as well as ray-traced $\Lambda$CDM $N$-body simulations, to measure the 3-dimensional cosmic shear induced by an abundance of clustering along the line of sight. We develop and test a method to reconstruct the foreground lens clustering by representing observed lensing peaks in the E-mode maps as clusters in a forward-model of the lensing, approximating the observed shear as a sum of successive lensing potentials. We begin in Section \ref{sec:massmap_method} with a description of the mass mapping method demonstrated on the Buzzard $N$-body simulations. In Section \ref{sec:simlens} we present our method of lens reconstruction on these simulations and compare the estimated E- and B-mode maps and correlation functions.  In Section \ref{sec:data} we introduce multi-color Suprime-cam \& {\it Hubble Space Telescope} observations of a deep weak lensing field in which we will apply this modeling technique. We briefly describe our data reduction process, which goes from raw data to photometric redshift calibration and shear measurement using the `stack-fit' algorithm on multiple dithered exposures (and multiple bands). With these calibrated shape and redshifts, in Section \ref{sec:simlynx} we apply the lens modeling technique to the Lynx field observations and demonstrate that these modeled halos capture most of the information in the E- and B-mode maps and correlation functions. We discuss these results in Section \ref{sec:discussion} and conclude in Section \ref{sec:summary} with a summary and future applications of the method to wide field surveys.

\section{Mass mapping method}
\label{sec:massmap_method}
In theories and $N$-body simulations of the formation of large scale structure, the clumping of matter forms in a hierarchical manner with well-defined and characteristic mass density profile for clusters of galaxies \citep{NFW1997ApJ}. Through weak gravitational lensing, each clump of galaxies similarly induces a characteristic pattern of shear on background galaxies \citep{Wright2000ApJ}. Therefore, one can use the measured weak lensing shear to infer the mass map, reconstructing the lensing field by optimally filtering the shear with a profile matched to the lensing signal of an NFW mass profile. We use this characteristic profile to apodize the measured shear field and locate overdensities along the line of sight by spatially mapping the aperture mass statistic $M_{ap}$:
\begin{equation}
  \label{eq:m_ap}
  M_{ap}(\theta)=\int Q(|\theta|)\gamma_t(\theta)d^2\theta
\end{equation}
as introduced in \citep{Schneider1996MNRAS}. This function takes the background galaxy shears around each point in a map and filters them with the function radial chosen function $Q(|\theta|)$. In our case, $Q$ is chosen to approximate the tangential shear profile of an NFW-like overdensity through an apodized combination of exponentials and hyperbolic tangent functions which can be scaled to match clusters of varying mass and lensing efficiency. This useful approximately-NFW filter function was introduced in \citet{Schirmer2007A&A}, and is given by Equation \ref{eq:qfilt}.
\begin{equation}
  \label{eq:qfilt}
  Q_{NFW}(x)=Q_{box}(x) \frac{tanh(x/x_c)}{x/x_c}
\end{equation}
The parameter $x$ is a dimensionless radius: the angular separation in units of the cutoff radius $r_{out}$, i.e. $x:=r/r_{out}$. The rate of tangential shear decrease (concentration) in the NFW profile $x_c$ is taken to be $x_c\sim 0.15$ for all map-making processes. The apodization function $Q_{box}$ provides the exponential damping around zero radius and around the cutoff radius $r_{out}$ and is given by:
\begin{equation}
  \label{eq:qbox}
  Q_{box}(r)=(1+e^{6-150r/r_{out}}+e^{-47+50r/r_{out}})^{-1}
\end{equation}
Maps using both this NFW filter and a generic polynomial filter are generated and compared, and both the E- and B-modes are found to be proportional. In the following analysis, only the maps with NFW-matched filter $Q_{NFW}$ are used as they have higher signal-to-noise contrast in both the simulations and observations.

Due to the WL mass-sheet degeneracy, convergence or mass density maps are measured by subtracting a reference shear from the measured shear, and thus are bipolar. Such maps are often thresholded at some positive level in order to display positive mass. Rather than picking thresholds or contours arbitrarily, we choose instead to plot aperture mass signal-to-noise ratio. Significance of mass density peaks may then be readily assessed.  We spatially map the aperture mass signal-to-noise $S_t$ by dividing $M_{ap}$ by the correspondingly apodized shape shot noise with an average RMS source galaxy shear of $\sigma_e = 0.3$. This denominator is equivalent to computing the statistic in a field with no lensing. This can be represented discretely as a sum over all background galaxies:
\begin{equation}
  \label{eq:s_t}
  S_t=\frac{\sqrt{2}\sum\epsilon_{t,i} Q_i}
	 {(\sum(\sigma_e Q_i)^2)^{1/2}},
\end{equation}
where $e_{t,i}$ is the galaxy's tangential ellipticity relative to the map pixel, $Q_i$ is the NFW filter weight at the galaxy's radial position, and $\sigma_e$ is the galaxy shape shot noise. Peaks in this $S_t$ map correspond to locations where the tangential shear is largest and most similar to an NFW mass profile, which most likely correspond to clusters or groups of galaxies in the foreground. Therefore we will use these shear peaks to estimate the location and mass of foreground cluster lenses.

Similarly, one can use the cross shear $\gamma_\times$, instead of the tangential shear in Equation \ref{eq:m_ap} to define the cross statistic:
\begin{equation}
  \label{eq:s_x}
  S_\times=\frac{\sqrt{2}\sum\epsilon_{x,i} Q_i}
	 {(\sum(\sigma_e Q_i)^2)^{1/2}}
\end{equation}
Unlike the maps of $S_t$, the $Q_{NFW}$ filter function is not necessarily optimal for B-mode detection, because there is no expected cross shear for a single isolated NFW concentration lensing a well-sampled and homogeneous background. Nonetheless, this B-mode statistic will pick up any axisymmetric curl-like shear pattern and can be easily computed alongside the E-modes. Often used as a check for systematic errors, significant B-modes in weak lensing studies are usually attributed to PSF mis-estimation or spatially varying calibration errors under the assumption that if PSF correction is done properly (and there are no astrophysical sources of B-modes), these maps should be consistent with noise. However, mapping itself can induce leakage from E-modes into observational B-modes in a predictable fashion for a given distribution of sources and lenses, and crucially, astrophysical B-modes can also be induced by multiple lensing and lensing of non-random source ellipticity distributions such as those induced by intrinsic alignments. Therefore both E- and B-mode maps are worth careful investigation in simulations and observations.

\subsection{Mapping N-body simulations}
We first investigate the mass mapping method on simulations using the Buzzard cosmological $N$-body simulations \citep{DeRose2018} that were produced for the Dark Energy Survey \citep{DES2005astro.ph} and which provide realistic 5-year depth galaxy shear and halo catalogs. Catalog data used include positions (RA/Dec/z) and reduced shear ($g_1/g_2$) of galaxies computed using the ray-tracing code $\texttt{CALCLENS}$ \citep{Becker2013PhDT} as well as the positions and masses $M_{200}$ of clusters in the halo catalog. After mass mapping with the filters described in Equation \ref{eq:qfilt} applied to Equation \ref{eq:s_t}, we can identify peaks in the mass map and correlate them with the (unobservable) halo catalog to develop our method of shear reconstruction.

\begin{figure}
  \includegraphics[width=\columnwidth]{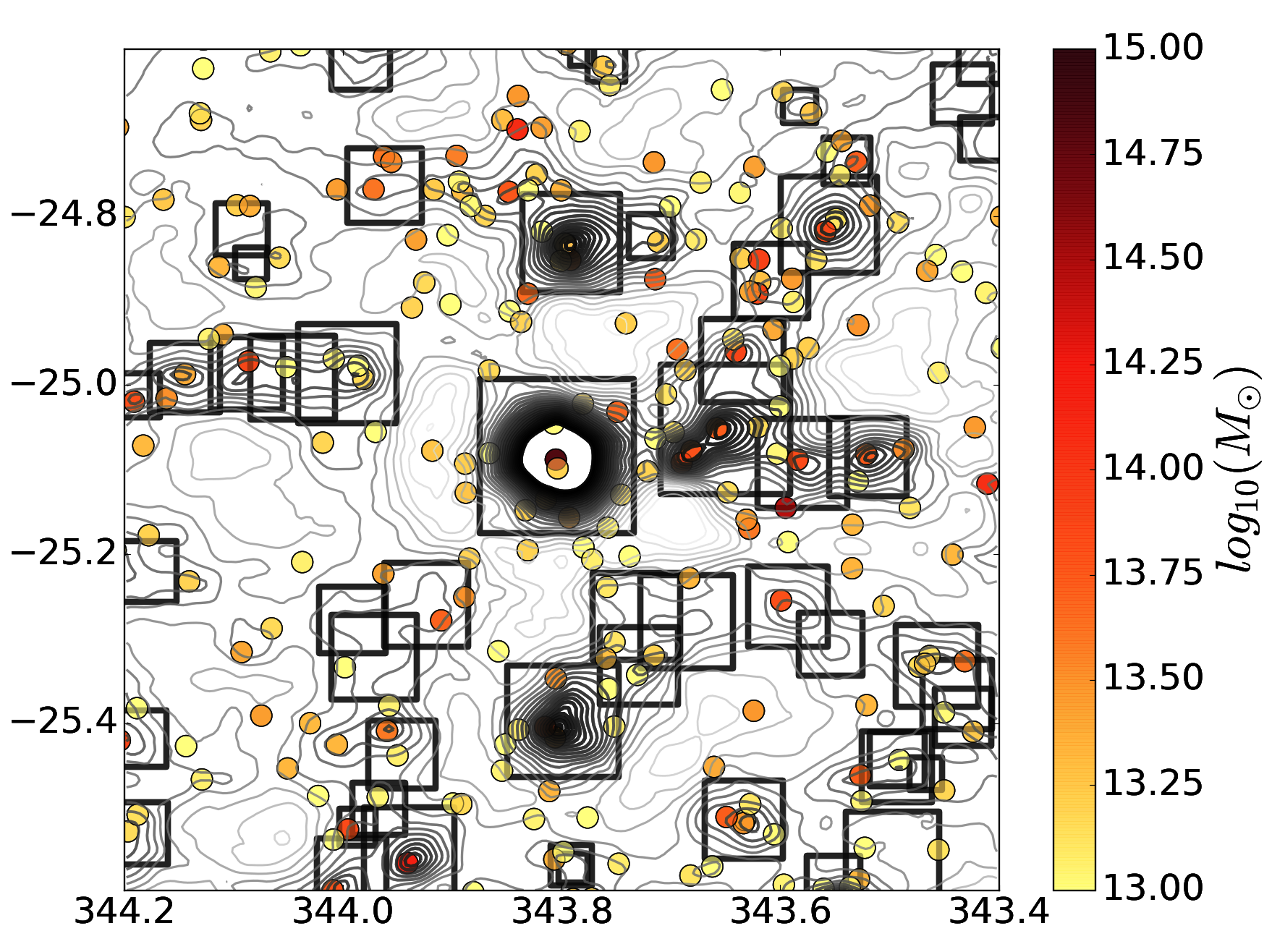}
  \caption{An E-mode aperture mass map, with grey-to-black contours showing $-2<S_t<5$ (Eq. \ref{eq:s_t}), produced by filtering N-body ray-tracing shears with an NFW-like filter (Eq. \ref{eq:qfilt}). The locations of massive ($M_{200}>10^{13}M_{\odot}$) halo catalog members are shown as dots with color proportional to their mass. The boxes show the locations of detected $S_t$ peaks which are used in the mass estimation and reconstruction of the halo catalog.}
  \label{fig:buzzard-detect}
\end{figure}

We choose to analyze a small region of the Buzzard universe $\sim 1^\circ\times1^\circ$ wide and centered around a massive foreground lens at ($\alpha$: 343.7, $\delta$: -25.0). The large cluster centered in the foreground of this slice is at redshift $z=0.28$ with mass $M_{200}=6.8\times 10^{14} M_{\odot}$ provides an anchor for our observations as it is easily detected given $\sim40,000$ background galaxies with an observed number density of $n_{gal}=13$ background galaxies per square arcminute. Additionally, there are $\sim$120 other clusters/groups with mass $M_{200}>2\times 10^{13}M_{\odot}$ which will also be used into the lensing simulation, shown as colored dots in Figure \ref{fig:buzzard-detect}. If we then slice the galaxy catalog by choosing galaxies only with $z_{photo}>0.5$, we can detect the collective imprint of these overdensities as the high S/N regions of the E-mode map shown as contours in Figure \ref{fig:buzzard-detect}. There is a clear correlation of the contours and overdensities of halo catalog members along the line of sight, confirming the obvious utility of mass mapping. In the next section we develop a method to model each E-mode peak as a lensing mass detect peaks using an image segmentation algorithm similar to that used in \texttt{SExtractor}. The locations of high aperture mass are shown in Figure \ref{fig:buzzard-detect} as black boxes.

\section{Lens modeling of the E \& B modes}
\label{sec:simlens}
\begin{figure}
  \includegraphics[width=\columnwidth]{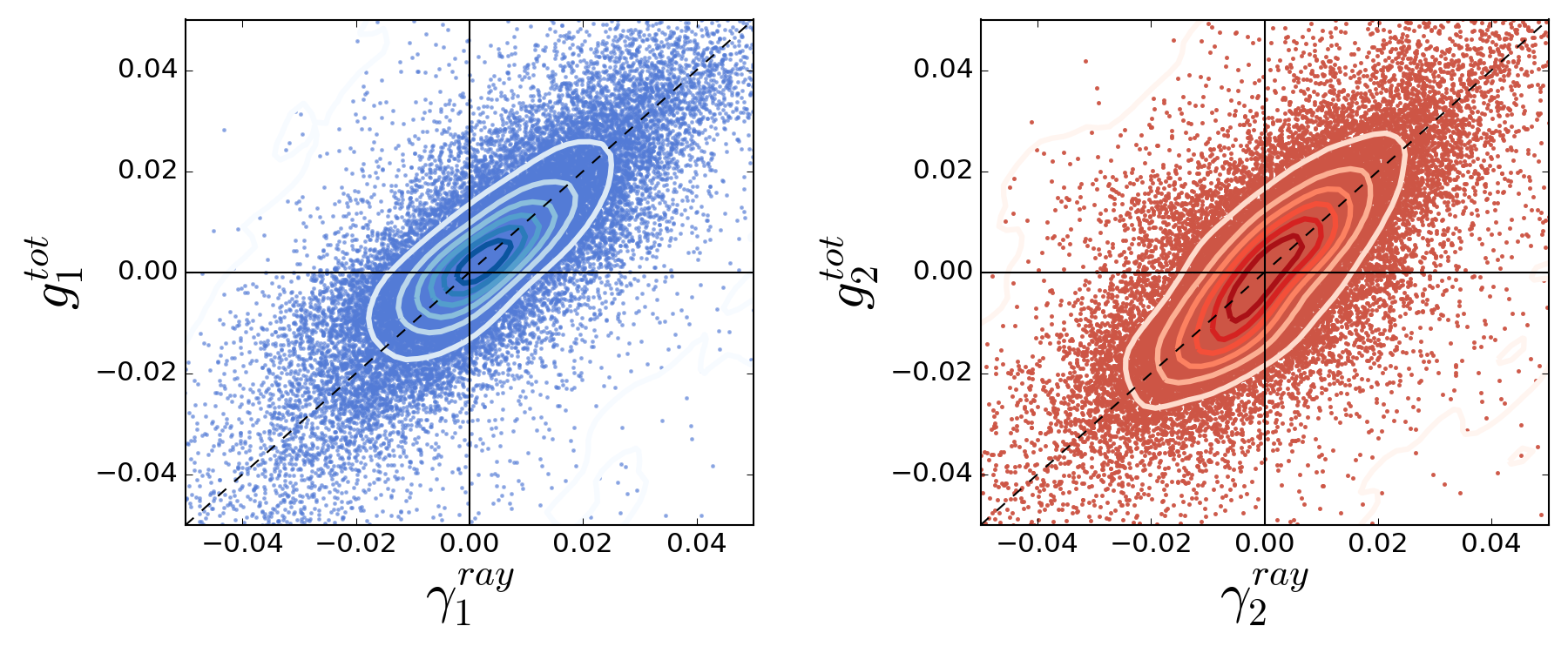}
  \caption{Comparison between the shear computed from ray-tracing an $N$-body simulation, $\gamma_{1,2}^{ray}$, and the reduced shears computed from the simple approximate method $g_{1,2}^{tot}$, summarized in Eq. \ref{eq:shear_tot}. There is broad agreement between both the components of the shear, confirming the usefulness of this method for approximating the ray-tracing process.}
  \label{fig:buzzard-e1e2}
\end{figure}

Our goal is to estimate the observed shear of background galaxies as the summation of the shears induced by the most massive clusters along the line of sight. In the weak lensing limit, the shears of each lensing halo add together, i.e. each background galaxy's total shear is represented as:
\begin{equation}
  \label{eq:shear_tot}
  g_{tot}=\sum\limits_{j}g_j(M_j,z_j,c_j,\delta_r,z_{source})
\end{equation}
where the sum is taken over all halos, $j$, which lie in the foreground of the galaxy, and where $M_j$, $z_j$, $c_j$, are the halo mass, concentration, and redshift, and $\delta_r$ and $z_{source}$ are the estimated angular distance between source and halo and the photometric redshift of the source. The resultant reduced shear sums for each galaxy in the simulation can then be compared to the observed shears. In the case of Buzzard, the observed shears are ray-traced using $\texttt{CALCLENS}$ \citep{Becker2013PhDT}, a computationally complex calculation. The simplification in Equation \ref{eq:shear_tot} can be tested in the weak lensing limit by using the positions and masses $M_{200}$ of clusters in the halo catalog, assuming an NFW shear profile \citep{Wright2000ApJ} and mass-concentration relation \citep{Duffy2008MNRAS}. Given a model of angular diameter distances between the lens and source (i.e. $h=0.7$,  $\Omega_\Lambda=0.7, \Omega_m =0.3$), each lens adds a weak shear to each source galaxy. We therefore approximate both components of the reduced shear of each source galaxy $g_{1,2}^{tot}$, which are robustly correlated to the ray-traced shears $\gamma_{1,2}^{ray}$ as shown in Figure \ref{fig:buzzard-e1e2}.

However, in an observational setting, we are not given a halo catalog with positions and masses. We must therefore estimate these unobservable `halo catalog parameters' from the data, namely through measurements of the reduced shear ($g_1/g_2$) and position (RA/Dec/z) of galaxies. Mass mapping provides the correlation between overdensities and high S/N regions of the $S_t$ maps, as illustrated in Figure \ref{fig:buzzard-detect}. We show the feasibility of applying this method to real data by estimating the locations and masses of halo catalog objects in the Buzzard universe, using only aperture mass maps. Overdensities are identified with peaks in the mass map (shown as black boxes in Figure \ref{fig:buzzard-detect}) using an image segmentation and deblending algorithm. These candidate cluster RA/Dec positions are clearly associated with objects in the (unobserved) halo catalog, shown as colored dots in that figure.

\begin{figure}
  \includegraphics[width=.9\columnwidth]{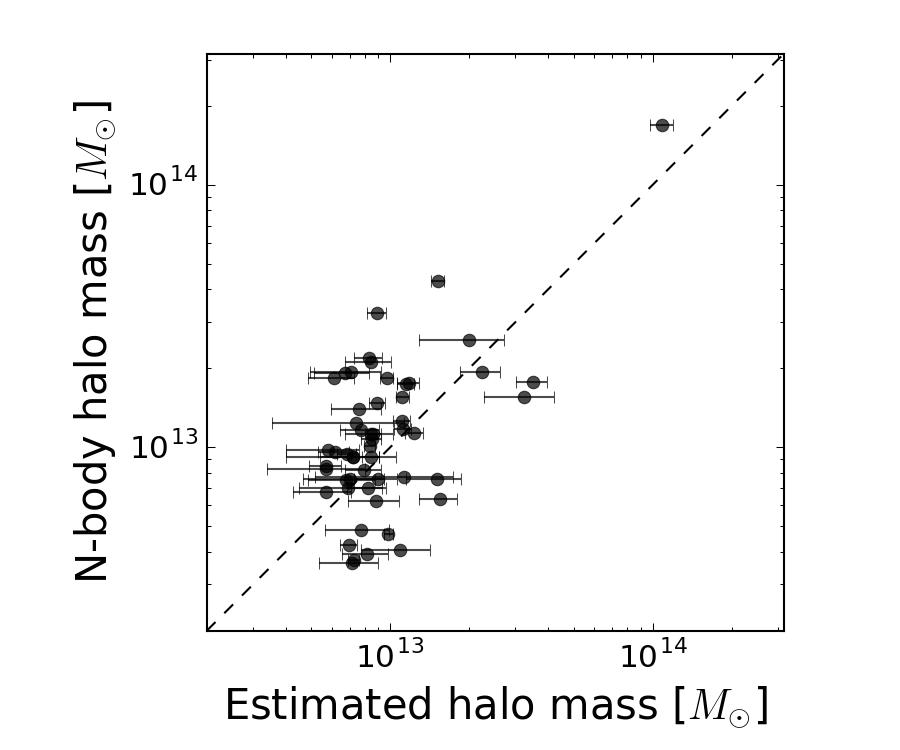}
  \caption{Approximating the masses in the halo catalog using only observational data. Lensing masses are estimated using the galaxy redshift catalog and fits $M_{ap}$ map characteristics such as peak value, integrated $S_t$, and width (Eq. 19 of \citet{Schneider2005astro.ph}). These few approximate lenses capture most of the shear information gathered from ray-tracing the full $N$-body simulation.}
  \label{fig:buzzard-masscorr}
\end{figure}

Given this list of peak RA/Dec positions, the mass and redshift of candidate clusters can then be estimated. Redshifts of clusters are estimated from the photometric redshift  distribution, $P(z_{phot})$, of foreground galaxies which lie along the line of sight to each peak, with the expectation that galaxies are tracers of the position of the overall halo. Instead of modeling each cluster halos individually, we simultaneously model all clusters using a combination of tangential shear profile fitting and correlations between aperture mass maps S/N and $M_{200}$ \citep{Schneider2005astro.ph}, which can be extrapolated from wide-field trends of the Buzzard simulation. The resulting correlation between this predicted mass and the halo catalog mass measured from the Buzzard $N$-body simulation is shown in Figure \ref{fig:buzzard-masscorr}. Because mass is the only source of lensing E-modes in the $N$-body simulations, these few approximate lenses capture most of the shear information from a much more complex and computationally intensive ray-tracing the full $N$-body simulation.

\begin{figure}
  \includegraphics[width=\columnwidth]{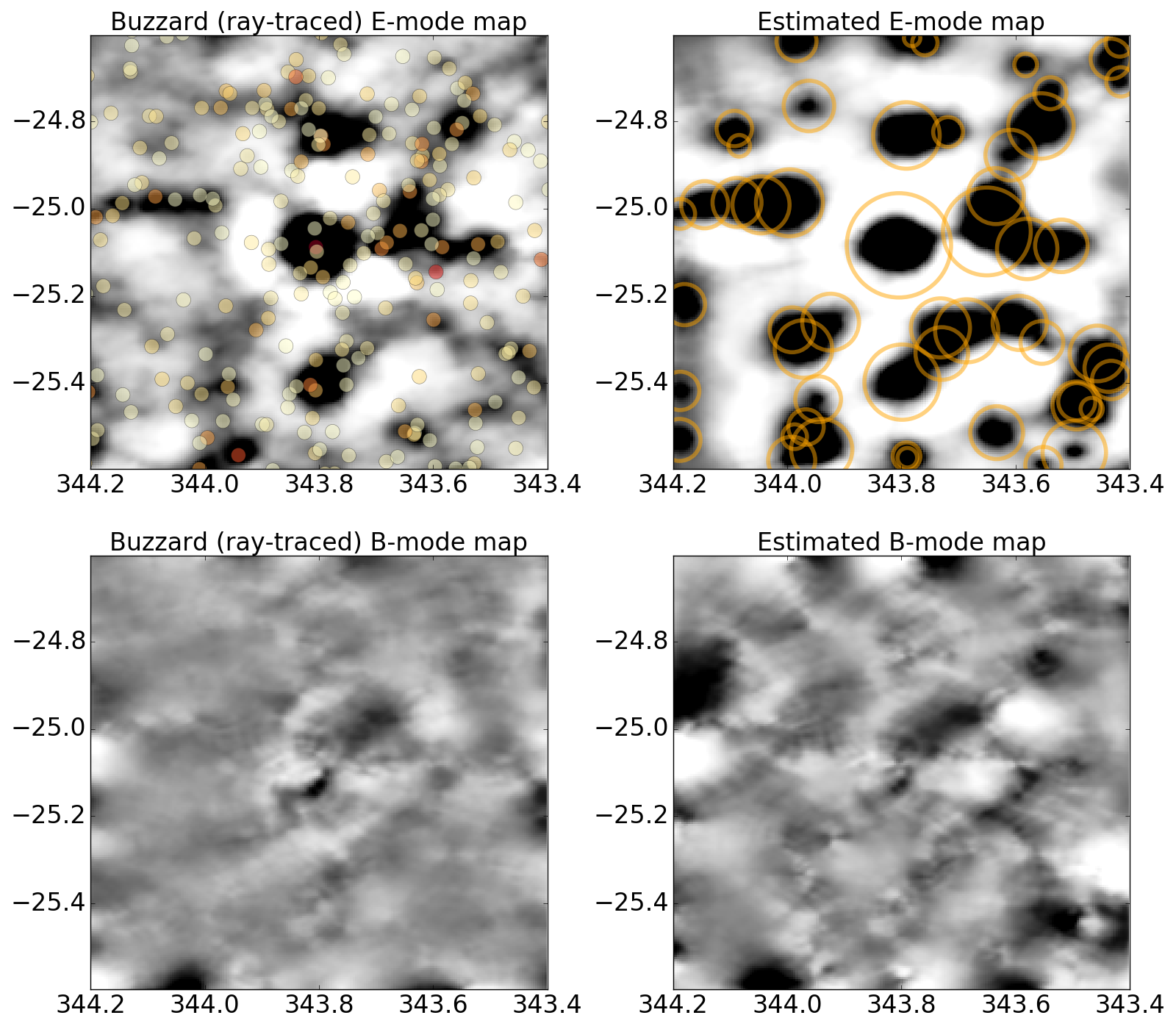}
  \caption{Comparison of E/B maps between Buzzard ray-traced shears and our estimated shears, all color bars are -1<$S_t$<1 white to black. In the left column, shear is calculated through application of \texttt{CALCLENS} multiple-plane ray-tracing algorithm through the Buzzard $N$-body simulation. Overlaid are the locations of halo catalog members color coded according to their mass as in Figure \ref{fig:buzzard-detect}. On the right, maps are computed in the same way but using shear estimated by summing the tangential shear around cluster masses estimated in Figure \ref{fig:buzzard-masscorr} at the locations indicated by open orange circles. In both ray-traced shears and our estimated ones, B-modes are present at similar amplitudes and locations. These B-modes must therefore be a result of multiple lensing and mass mapping artifacts, including source clustering and edge effects.}
  \label{fig:buzzard-ebmap_compare}
\end{figure}

Using the cluster mass and location estimates we then simulate the shear of the background galaxies as the sum of the predicted cluster mass lensings, thus introducing only E-modes into the shear field, according to Equation \ref{eq:shear_tot}. Application of Eqs. \ref{eq:s_t} and \ref{eq:s_x} on these shears produces estimated aperture mass maps $S_t$ and $S_\times$  (E and B-mode) maps which we can then compare to the ones which utilize ray-tracing shears. As shown in Figure \ref{fig:buzzard-ebmap_compare}, the overall pattern in both modes is quite similar between ray-traced and approximated maps. The E-modes in particular are very strongly similar, as our simulated foreground lenses have been chosen to match the observed tangential signature maps. However, there is much more small scale structure in the ray-traced shears which are estimated from the full matter power spectrum of the $N$-body simulations. For the B-mode, the simulated and observed maps also have statistically significant correlations, which we can explore in several scenarios. 

\begin{figure}\centering
  \includegraphics[width=\columnwidth]{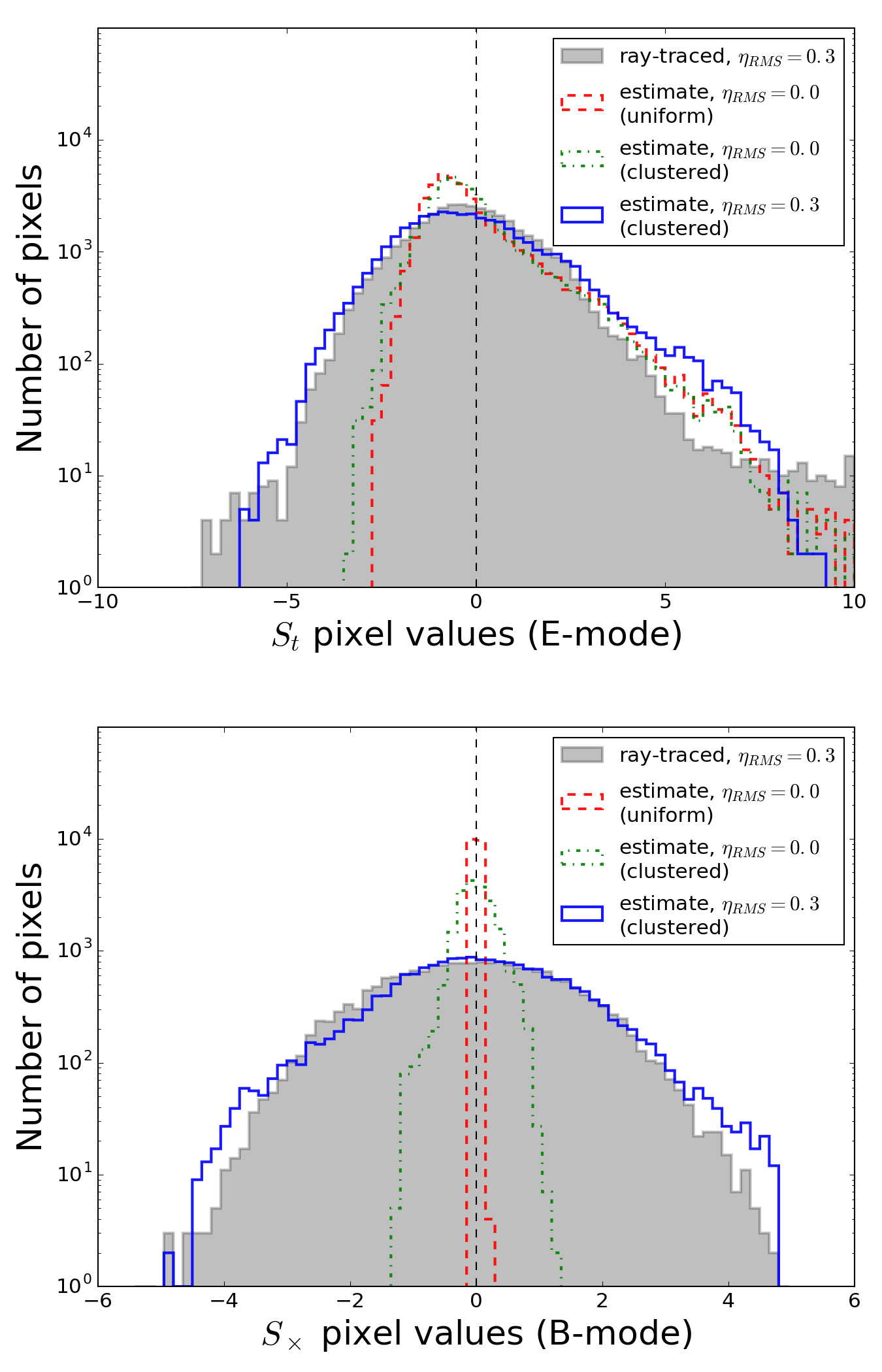}
  \caption{Comparison of our ray-tracing approximation method, using the distribution of E- and B-mode map pixel values in a sub-field the Buzzard $N$-body simulation (shown with logarithmic vertical scaling). The approximate lensing model used here is informed from only observable quantities, and has no knowledge of the halo catalog. Red lines show the modeled shears of a uniform (homogeneous) source plane of galaxies at $z=1.0$ that fill the observed field of view. Green lines use the actual (inhomogeneous) distribution of source galaxies. Blue lines show the distributions after the addition of shape noise, which better matches the observed distributions (shaded). B-mode distributions are shown without edge effects, indicating the presence of multiple lensing B-modes in the uniform galaxy simulation.}
  \label{fig:buzz-ebhist}
\end{figure}

The first is the addition of shape noise, which smooths the large scale structure of the maps adds noise to the small scales, thereby broadening the distribution of $S_t$ and $S_\times$ pixel values as shown in the histograms of Figure \ref{fig:buzz-ebhist}. It is shown that if random shape noise is included in the forward simulation at an amplitude of $\eta_{rms}=0.3$, it is found that that the 1-$\sigma$ width of the simulated $S_t$ and $S_\times$ distributions nearly match the width of the observed $S_t$ and $S_\times$ distributions as seen in Figure \ref{fig:buzz-ebhist}. At high ($S_t>5$) values, a slight over- and then under-shoot in the distribution of E-mode map pixels can be seen. These small ($\sim 1\%$) discrepancies indicate that more precise modeling of the halos is needed to account for all pixel values. For instance, the empirical mass-concentration relation assumed in the approximation of the halo lensing potential \citep{Duffy2008MNRAS} may need to be adapted to the specifics of each N-body simulation.

Beyond those induced by shape noise, there are other sources of observational B-modes at play here. The largest and most mundane source of B-modes are those induced at the boundaries of the aperture mass map, where pure E-modes can leak into the B-mode due to azimuthal symmetry violation in aperture measures. These B-modes can be mitigated by padding the edges of the simulation or by limiting consideration of B-modes near the edges entirely, and so they are excluded from the analysis shown in Figure \ref{fig:buzz-ebhist}. However, these edge B-modes are related to the strengths and locations of the lenses in the field, and therefore are not entirely noise. Another source of B-modes is the clustering of source galaxies, as discussed in \citet{Schneider2002A&A} and \citet{Yu2015ApJ}. These variations in spatial positions of galaxies again lead to an asymmetry in aperture measurements of pure E-mode fields, and again are completely describable if the known positions of E-modes are known.

We can probe and eliminate these two galaxy density effects in our simulation by replacing the realistically clustered  galaxy distribution in RA/Dec/z with a uniform distribution of galaxies at a fixed redshift but which are lensed by the estimated 3D positions of halos. This has the effect of drastically reducing the observed B-modes, indicating that the majority of B-modes in mapped fields results from the realistic source density variations. Additionally, If we map the E- and B-modes using a uniform distribution of galaxies (at a single source plane of $z=1.0$) extended beyond the edge of our slice, such that no cluster is within several aperture radii of the extended galaxy boundary, the edge effect disappears. However, as shown in Figure \ref{fig:buzz-ebhist}, there are still non-zero B-modes even in the case of uniform source galaxy distributions with no edge effects.

B-modes which remain after edges and source non-uniformity effects are removed may prove to be physically interesting. For instance, they can be a sign of multiple lensing of background galaxies, which produces a characteristic quadrupole-like B-mode pattern when two lenses are well-aligned along the line of sight \citet{Bertin2001ApJ}. However, the simulated source galaxy density in that study is much higher than our current ground observations allow, and their simulations only include very fortuitous alignments of clusters along the line of sight. Therefore, we do not expect such obvious B-mode patterns in our observations, and they aren't readily seen in our observed maps. However, in our simulations we can probe this multi-lensing effect by modifying the distribution of lenses and galaxies. In fact, it is observed that the width of the $S_\times$ distribution is broader in the case of realistic 3D lens positions than when all lenses are placed at a single redshift. Additionally, if the source galaxy density is greatly increased to $n_{gal}=100 ~ \rm{arcmin}^{-2}$, the width of the $S_\times$ distribution is increased even further and the quadrupolar B-mode pattern of double lensing is observable in the maps. In practice, this multi-lens effect broadens the simulated (and observed) $S_\times$ distribution by a small amount, but given the much lower surface density of $z>0.8$ galaxies in our observations, the distinct pattern of double lensing in this field is unresolvable.

\subsection{Shear correlation functions}
\label{subsec:buzzard-shear_corr}
Statistical correlations are complimentary to spatial maps. While losing spatial information, the summary statistics produced by correlating the shears of all galaxies can provide a less noisy test of the origins of B-mode. Correlation functions over the field are computed using both the ray-traced shears and those simulated under multiple conditions. The shear correlations $\xi_{\pm}$,  the E- and B-mode aperture mass dispersion $\langle M_{ap,\times}^2 \rangle$, and top-hat shear $\langle \gamma^2 \rangle_{E,B}$ dispersion are calculated on scales of $0.5<\theta<15$ arcminutes. The raw shear correlation function is computed by correlating the tangential and cross shear around each galaxy, and either summing or differencing the two correlations.
\begin{equation}
  \label{eq:xipm}
  	\xi_{\pm}(\theta) = \langle \gamma_t\gamma_t\rangle \pm \langle \gamma_\times\gamma_\times\rangle 
\end{equation}
These raw correlations depend on both the E- and B-mode power spectrum, and so the decomposition into aperture mass statistics is a useful one. We define the aperture mass correlations in terms of the $\xi_{\pm}$ functions generally as:

\begin{equation}
  \label{eq:mapcorr}
  	\langle M_{ap,\times}^2 \rangle(R) = \int_{0}^{\infty} \frac{r dr}{2R^2} \left [ T_+\left(\frac{r}{R}\right) \xi_+(r) \pm T_-\left(\frac{r}{R}\right) \xi_-(r) \right] 
\end{equation}
where $M_{ap}$ and $M_\times$ take the plus and minus signs, respectively, and the functions $T_{\pm}$ are window functions which are generalized autocorrelations  of the filter function $Q$, and which limit the inclusion of small and large radii which are difficult to measure. Previously in Section \ref{sec:massmap_method} we chose $Q$ to be a signal-matched filter (NFW-like) for optimal detection of E-modes, but for the correlation functions presented here we use a Gaussian-derivative  type window function which give $T_\pm$ in the form:
\begin{eqnarray}
\label{eq:tpm}
    T_+(s) =& \frac{s^4 - 16s^2 + 32}{128} \exp(-s^2/4) \\
    T_-(s) =& \frac{s^4}{128} \exp(-s^2/4)
\end{eqnarray}
as presented in \cite{Crittenden2002ApJ}. 

\begin{figure}
  \includegraphics[width=\columnwidth]{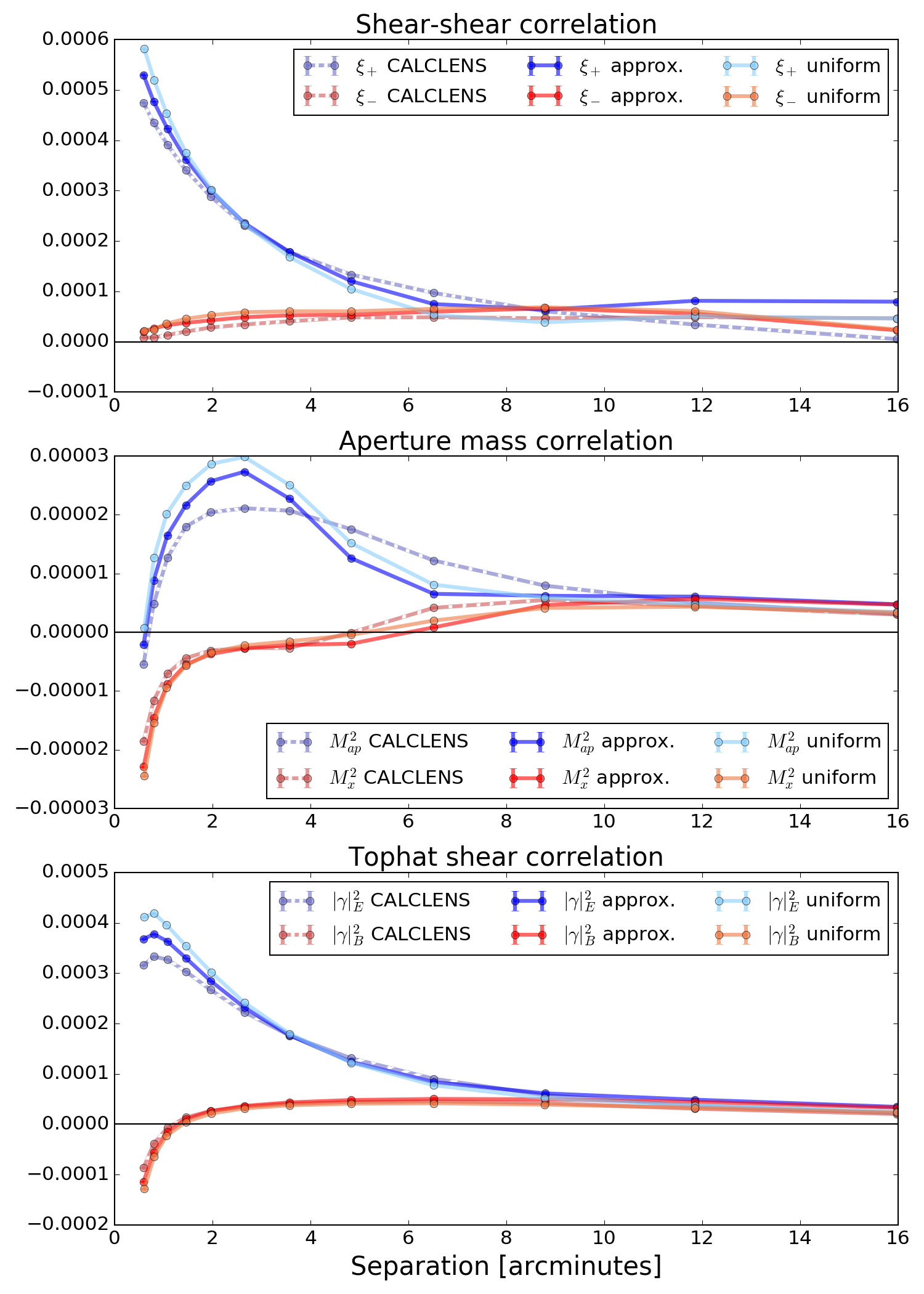}
  \caption{The three shear correlation functions for the estimated shears as compared to those from N-body ray-tracing. Shear-shear $\xi_{\pm}$, aperture mass $\langle M_{ap,\times}^2\rangle$, and top-hat shear $\langle \gamma_{E,B}^2\rangle$ are shown in the top, middle, and bottom rows and defined in Eqs. \ref{eq:xipm}, \ref{eq:mapcorr}, and \ref{eq:gamsqcorr} respectively. In each panel the spatial correlations (both E- and B-modes) of the estimated and N-body ray-traced shears can be compared and found to be similar in amplitude, shape, and scale.}
  \label{fig:buzzard-correlations}
\end{figure}

Another popular cosmic shear statistic is the shear dispersion within a circle of radius $R$, which can again be decomposed into the shear-shear correlations as given by the expression:
\begin{equation}
  \label{eq:gamsqcorr}
     \langle \gamma^2 \rangle_{E,B}(R) = \int_0^{2R} \frac{r dr}{2 R^2}
    \left [ S_+\left(\frac{r}{R}\right) \xi_+(r) \pm
    S_-\left(\frac{r}{R}\right) \xi_-(r) \right ] 
\end{equation}
where the E- and B-modes correspond to the $+$ and $-$ in the equation on the R.H.S. These E/B decompositions of the shear field are also filtered analogously to $\langle M_{ap,\times}^2\rangle$, but with the $S_{\pm}$ window functions applied as follows:
\begin{eqnarray}
	\label{eq:spm}
     S_+(s) =&\frac{1}{\pi} \left(4 \arccos(s/2) - s \sqrt{4-s^2} \right) \\
     S_-(s) =&\frac{1}{\pi s^4}[s \sqrt{4-s^2} (6-s^2) - 8(3-s^2) \arcsin(\frac{s}{2})] 
\end{eqnarray}
for $s<=2$, and $S_-(s) =  4(s^2-3)/(s^4)$ for $s>=2$. These window functions are broader than $T_\pm$, implying that more shear dispersion signal is included but the signal is less localized.  A constant shear generates contributions to both E- and B-modes \citep{Schneider2002A&A}. The correlations are computed on our dataset using the \texttt{TreeCorr} code \citep{Jarvis2004MNRAS}, and the results of the calculations for are shown in Figure \ref{fig:buzzard-correlations} using three sets of shears which have similar shape and scale.

The three separate shear datasets are as follows: the shears from the ray-tracing code $\texttt{CALCLENS}$, those simulated using the observed $RA/Dec/z$ positions clustered on the sky and lensed by our approximate foreground, and those simulated with uniform source plane at $z=1.0$ at a galaxy number density of $13 ~ \rm{arcmin}^{-2}$ lensed by the same foreground model. E- modes for each statistic are shown in blue shades and B-modes are shown in red shades. Error bars on the correlation functions are from variance in the shear itself, the simulated cases have no shape noise added. As a null test, randomizing the shear components but keeping the positions fixed resulted in zero correlation in all statistics at all scales. The slight mismatch (overshoot/undershoot) in the approximate model correlation functions indicate some residual tuning may be required. For instance, the empirical mass-concentration relation we assume \citep{Duffy2008MNRAS} is likely an improper approximation when comparing to ray-traced N-body simulations as ray-tracing accounts for mass at scales which are inaccessible to actual observations. Indeed, modification of the concentration relation does lead to a difference in the shape of the correlation function at small scales. Despite the fact that these two-point correlation functions were not fit to the data, the general agreement in the shape and scale of the correlations strongly suggests that this method of reconstructing halos can provide physical insight into both the E- and B-modes.

For instance, this field has extensive large scale structures which give an average matter density in excess of that given by the universal mass power spectrum. Given this clustering and depth of the observation along the line of sight, it is thus expected that the measured correlation functions would reach large values in agreement with the model. Interestingly, there are also non-zero B-mode in the measured correlations. Ordinarily these might be blamed on incomplete PSF modeling or gaps in the data, however there is no PSF in the simulations nor are their bright star masks or other depth variations. Though there were B-modes induced at the boundaries of the mass maps due to incomplete sampling of E-modes, in the case of these correlation functions the edges are not an issue. This is because $\xi_{\pm}$ are directly sampled at each galaxy point, and not on a grid as in the mapping scenario \citep{Schneider2002A&A}. In fact, masking more ($\sim 5$ arcmin) of the edges of the observation does not noticeably change the correlation function. Additionally, the ray-traced and approximated B-modes are of similar amplitude even when there is no masking and a uniform lens plane is used. This implies that the majority of observed B-modes in these correlation functions are not due to the PSF (which is not included in the simulations) or source galaxy clustering. Rather, the B-modes in the correlation function must be due to multiple lensing and artifacts intrinsic to the particular decompositions of a realistic shear field into E- and B-modes. In fact, \citet{Kilbinger2006A&A} show that E- and B-modes can mix when on small scales when there is a lack of close projected pairs of galaxies, and on large scales due to the finite field size. Both of these unavoidable observational facts limit complete sampling of the shear power spectrum and are present in all maps and correlations, to a degree determined by the depth of the field and the distribution of mass within it. Therefore, careful modeling of the lensing field can turn these B-modes into tools for the mapping of large scale structure, and understanding of these observational artifacts is also critical if astrophysical sources of B-modes are to be sought. As seen in the maps, histograms, and correlation functions of Figures \ref{fig:buzzard-ebmap_compare}, \ref{fig:buzz-ebhist}, and \ref{fig:buzzard-correlations}, these expected sources of B-modes are well captured by our lensing approximation and justify its use on observational data, as we do in the following section.

\section{Application to deep weak lensing observations}
\label{sec:data}
\begin{figure}
  \includegraphics[width=\columnwidth]{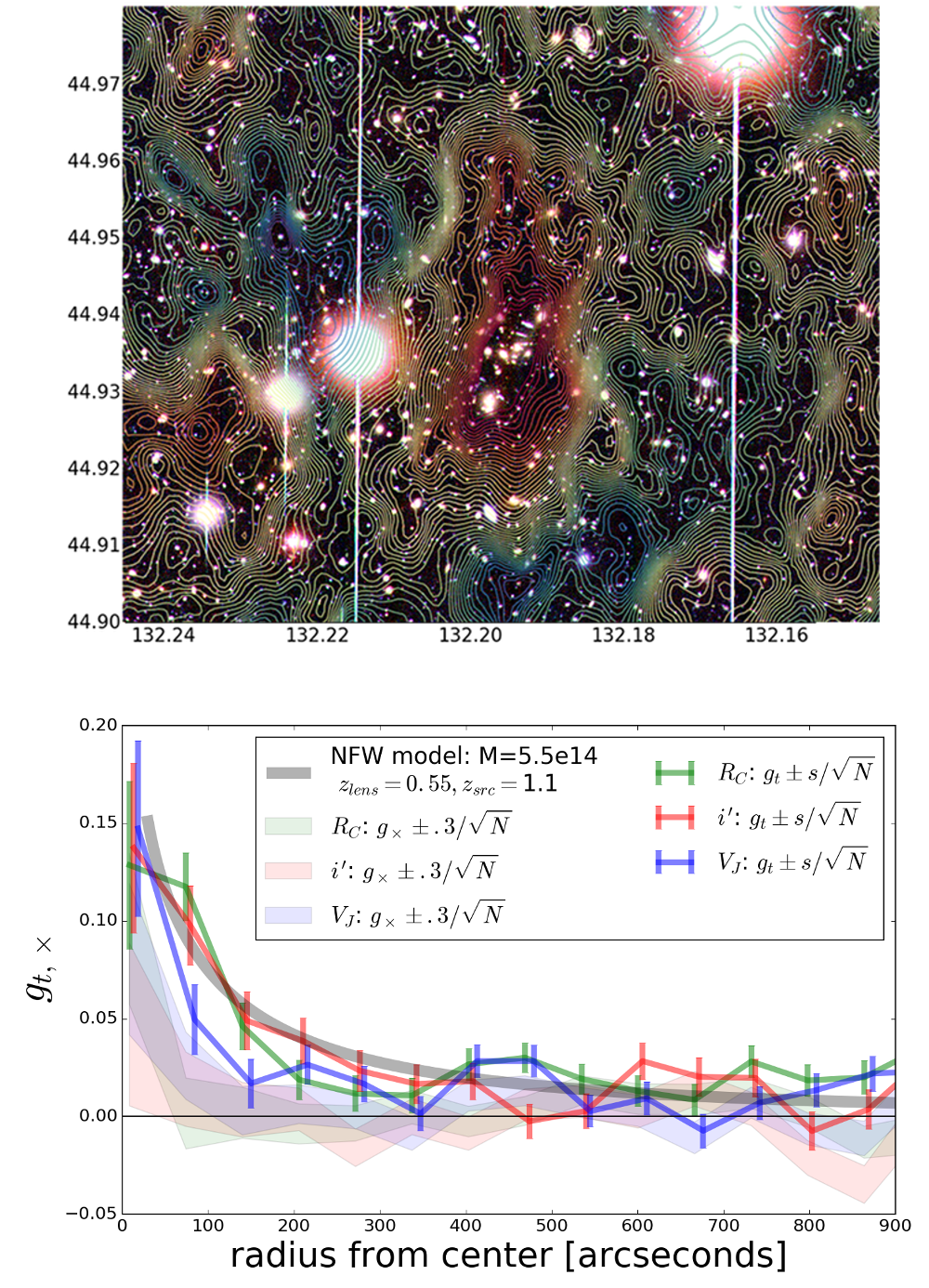}
  \caption{Lynx North, with mass $M=5 \times 10^{14} M_{\odot}$ at $z=0.55$, seen in imagery and measured shear signal. Top: a 5 arcminute wide subfield image of Lynx North. Mass map contours have been overlaid, ranging from $S_t$ of -3 to 8 as blue to red, where $S_t$ is defined in Equation \ref{eq:s_t}. The images in the $i', R_c,$ and $V$ band data provide the RGB colors. Structures of foreground red cluster sequence galaxies are often observed to correlate with the E-mode shear signal of background galaxies (those with $z_{phot}>0.8$). Bottom: the reduced shear measured azimuthally around Lynx North, as measured in the $V/R_c/i'$ wavelengths for sources with $z_{phot}>0.8$ and median redshift $z_{phot}=1.1$. A model of the tangential shear of a $M=5 \times 10^{14} M_{\odot}$ cluster induced on lensed sources at a median redshift of $z=1.1$ is shown in black. Tangential shears are shown as lines with error bars from the sample standard deviation in each bin ($s/\sqrt{N}$. Comparatively, cross shear measurements are shown as filled color curves, of width $.3/\sqrt{N}$ due to assumed intrinsic shape noise. The cross shear profile roughly agrees between the bands, showing an anomalous cross shear on the smallest scales which is not present in a singular NFW model. No bin overlap or smoothing has been applied.}
  \label{fig:lynx-north}
\end{figure}

The observational weak lensing data used in this study consists mainly of Suprime-cam 80 megapixel mosaic exposures in Lynx ($\alpha_{J2k}$: 132.20, $\delta_{J2k}$: +44.93) using the [$B/V/R_c/i'/z'$] filters for [60, 80, 90, 65, 90] minutes on the Subaru telescope during the first observing season in 2001 \& 2002. The exposures were dithered during observation, providing a uniform coverage of a field containing a large cluster, Lynx North (RX J0848+4456, mass $M=5\times10^{14} M_{\odot}$ and redshift $z=0.55$ \citep{Holden2001AJ}). Additional super-clustering of galaxies at redshift $z\sim 1.3$ has been discovered through a combination of X-ray surveys and galaxy clustering (\citet{Rosati1999AJ}, \citet{Stanford1997AJ}, \citet{Mei2012ApJ}). The Lynx field is similar to the simulated field in Section \ref{sec:simlens}; both fields are not the most dense regions of the Universe, but do feature a single large overdensity $\sim 10^{14} M_\odot$ and associated nearby large scale structure. We will use all detected galaxies behind $z>0.8$ (number density $17 ~ \rm{arcmin}^{-2}$) to map the abundance of foreground structure through the measured shear patterns they induce. A segment of the observations centered around Lynx North are shown in Figure \ref{fig:lynx-north}, where we show mass map contours overlaid on an RGB composite from $i', R_c,$ and $V$ band coadds frames in the top panel, and multi-color tangential shears measured around the cluster in comparison to an NFW model.

Images of the field are gathered from the Subaru-Mitaka-Okayama-Kiso Archive (SMOKA) system \citep{Baba2002ASPC}. Raw images were reduced for scientific analysis partially using the \texttt{SDFRED1} data reduction process, which provides Suprime-cam image overscan and bias subtraction, flat fielding, atmospheric dispersion correction, as well as masking of known bad pixels (\citet{Yagi2002AJ}, \citet{Ouchi2004ApJ}). Reduced images are then astrometrically aligned using \texttt{SCAMP} \citep{Bertin2006ASPC} by matching to known objects in the Sloan Digital Sky Survey (SDSS) catalogs. The analysis pipeline must fork at the step of image combination, because two different criteria must be optimized for weak lensing and photometric analysis which we describe separately.

\subsection{Photometric analysis}\label{subsec:phot}
For photometry, all images in a given $BVR_ci'z'$ filter are PSF matched and coadded using the \texttt{SDFRED1} pipeline. PSF matched photometry is then performed with \texttt{ColorPro} \citep{Coe2015ascl.soft}, where we form a cross-band detection image using the deepest and best seeing frames in the $R_c, i',z'$ (reddest) filters, weighted by their depth. This detection image is then degraded to individual filters where the seeing is poorer in order to estimate the isophotal flux lost by the degradation. PSF stars for this process are chosen using the same method described below in the shape measurement section \ref{subsec:shape}. Degradation of the images to a common seeing is performed using \texttt{IRAF}'s \texttt{psfmatch} kernel algorithm. This process results in a PSF-corrected magnitude that provides robust colors for faint and small galaxies which are close to the noise floor in images \citep{Coe2006AJ}. The resulting photometric catalog is then zero-point calibrated using a combination of matching to SDSS data as well as stellar locus regression \citep{High2009AJ} to determine precise absolute magnitudes across the coadded images. Besides providing foreground/background separation, the photometric color catalog is also useful in the identification of stars for PSF modeling in the following subsection. Counts of detections (3 or more pixels above $3\sigma$) indicate catalog completeness to $\sim 27$th magnitude, similar to the depth expected at the completion of the 10-year LSST survey. 

\begin{figure}
  \includegraphics[width=.9\columnwidth]{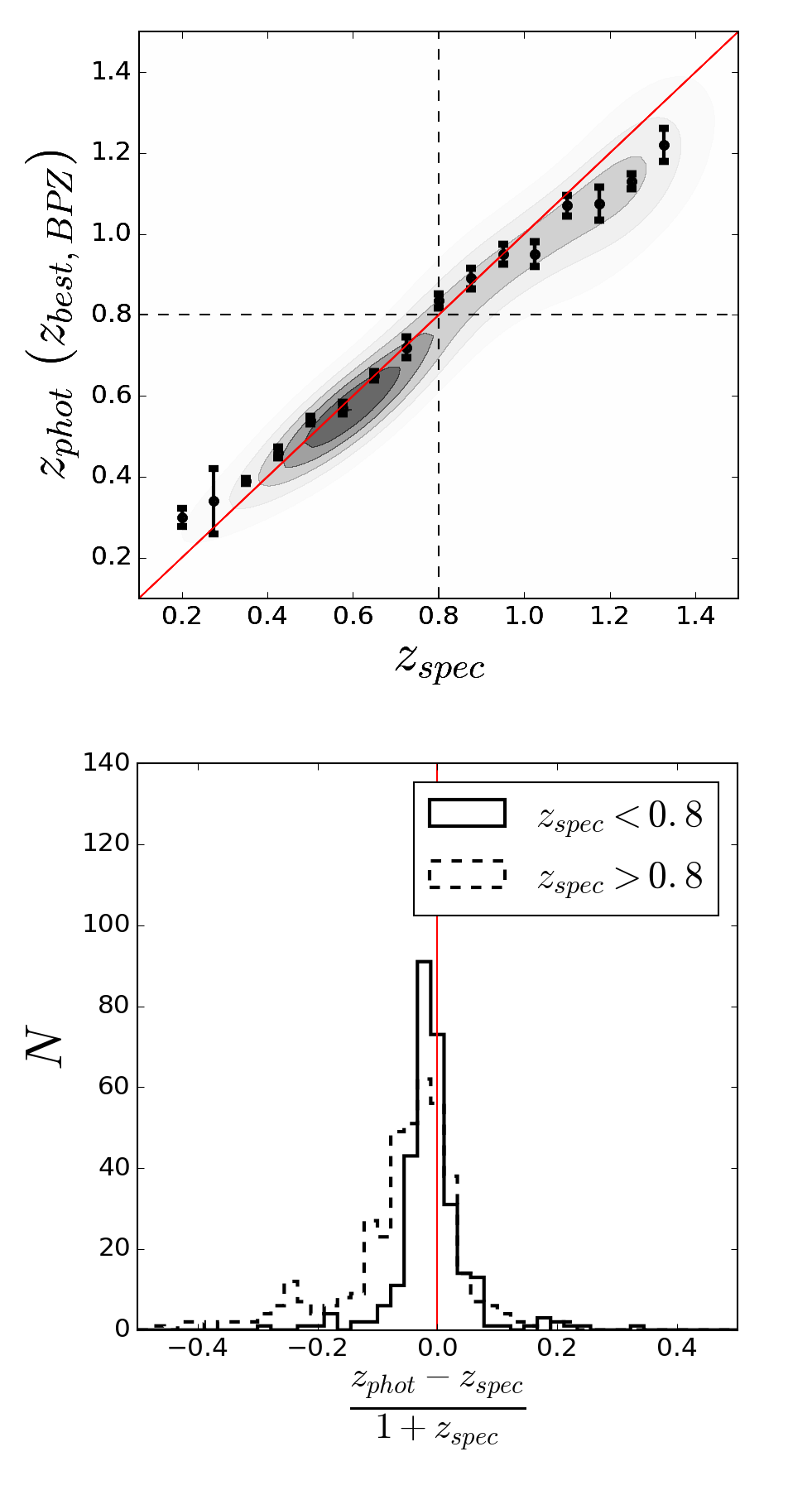}
  \caption{Photometric redshift calibration in the Lynx field observations. Top: Comparing hundreds of spectroscopically confirmed galaxy redshifts $z_{spec}$ to photometric redshift $z_{phot}$ (BPZ's $z_{best}$) estimates in redshift bins, where the error bars are the uncertainty on the mean of each bin and the dashed line indicates the $z>0.8$ redshift cut used in the mapping and correlation function analysis. Shaded contours show the redshift distribution of spectroscopic coverage. Bottom: Dispersion of redshift calibration was measured to be $\sigma_z\sim 0.056$  for $z<0.8$, implying distinct separation of foreground and background sources and allowing identification of galaxy clustering along the line of sight.}
  \label{fig:zphot_zspec}
\end{figure}

The calibrated broadband galaxy colors are then matched to models of redshifted galaxy spectral energy distributions (SEDs) using the Bayesian Photometric Redshifts (BPZ) software \citep{Benitez2000ApJ}, which provides estimates of the likelihood of redshift and type, $P(z,t)$, for every object. Some modifications to the defaults were made, to both improve the magnitude prior as well as to use updated galaxy SED templates which proved useful in the Deep Lens Survey \citep{Schmidt2013MNRAS}. These tweaks reduced the overall scatter and bias, ensuring separation of foreground and background galaxies critical to the weak lensing analysis. The photometric measurements in the field are further calibrated for zeropoint offsets using spectroscopic redshifts \citep{Mei2012ApJ} in the Lynx field. As seen in Figure \ref{fig:zphot_zspec} there is a clean separation between foreground and background galaxies at $z>0.8$, with $\sigma_z = \sigma[(z_p-z_s)/(1+z_s)]=0.084$ for all $z$ and $\sigma_z =0.056$ for galaxies $z<0.8$. This successful calibration shows that it is extremely unlikely for foreground sources to have scattered into the background sample, mitigating a large risk in projected and tomographic weak lensing measurements.

\subsection{Shape analysis}\label{subsec:shape}
Focusing on the weak lensing analysis, image quality must not be compromised via the above PSF equalization process which degrades images to match the PSF of the worst seeing filter. We instead consider each exposure and its PSF in a joint stack-fit for the galaxy shape which takes advantage of the best seeing exposures. The $V/R_c/i'$ images were determined to be of sufficient PSF quality for weak lensing by examining the exposures in each filter and measuring the mean width of the PSF and depth of imaging, which disqualified the $B$ and $z'$ filters respectively. In each of the $V/R/i'$ bands, non-PSF matched coadds are produced through \texttt{SWarp} \citep{Bertin2002ASPC}, after which each image of the coadd is individually and jointly analyzed. Preliminary object shape parameters are estimated in addition to the photometric measurements already computed on the coadd. Stars for PSF estimation are then selected through a combination of filtering and clustering in multi-dimensional parameter space. First, since our field overlaps with SDSS we are able to identify the positions of spectroscopically confirmed stellar objects, likely including some fraction of binaries unsuitable for PSF analysis. However, we can use these stars in each frame to find the location of the PSF-like objects in brightness/size space (a tilted line due to instrumental ``brighter-fatter" or charge transport effects \citet{Antilogus2014JInst}), which then allows for fainter non-spectroscopically confirmed PSF-like objects to be gathered. We then further require that these PSF-like objects occupy a $1\sigma$ region in color-color (all permutations of $BVR_ci'z'$) space occupied by the spectroscopically-confirmed objects. Finally, we also reject PSF outliers in $e_1-e_2$ PSF ellipticity space, and use the remaining stars as anchors for a model of the spatial variations of the PSF pattern in each exposure.

This observed PSF pattern of orientation and ellipticity is unique in each exposure, though it does show common smoothly-varying features which are indicative of previously investigated misalignment and drifting of optical elements during observation \citep{Hamana2013PASJ}. We model these field-wide aberrations using dozens of PSF stars on each CCD chip and hundreds in each exposure with a smoothly varying principal component analysis (PCA) model.  This PCA model finds the coefficients of twenty ``eigenPSFs'' which describe the majority of PSF variance in each exposure, and we then fit a polynomial surface to the PCA coefficients to provide a map of the PSF across the focal plane for each exposure, as in \citet{Jee2007PASP}. Special care has been given to ensuring the edges of each chip are roughly continuous and unaffected by gaps in PSF data.

Once the PSF stars have been selected, modeled, and interpolated to the position of each galaxy in an individual exposure, the process of forward PSF convolution and shape estimation of galaxy images can begin. The shape measurement algorithm used in this study operates on each exposure and is a modified form of the one used in the DLS \citep{Jee2016ApJ} called \texttt{sFIT} which won the GREAT3 gravitational lensing challenge \citep{Mandelbaum2015MNRAS}. For each galaxy the algorithm fits an elliptical Gaussian jointly across all images using the spatially resolved PSF extrapolated to the position of the galaxy in each exposure. Shear calibration for this method is provided through image simulations. Though precise and accurate shape measurement is an ongoing challenge to the weak lensing community, it can be seen that shear calibration is secondary to the large cosmic shear signal in this field. Additionally, the weighting applied by multiplicative bias calibration does not affect the value of $S_t$ because it changes both the numerator and denominator in Eq. \ref{eq:s_t} in similar ways \citep{Miyazaki2018PASJ}.

As one consistency check, we investigate the agreement between wavelength bands on the final measured ellipticity profile of Lynx North, shown as colored lines in the bottom panel of Figure \ref{fig:lynx-north}. A model of the tangential shear of background galaxies according to an NFW profile with mass, concentration, and redshift of $M=5\times10^{14} M_{\odot}$, $c=4$, and $z=0.55$ is also shown, measurements which have been confirmed through strong lens modeling and X-ray analysis of the galaxy cluster (\citet{Rosati1999AJ}, \citet{Stanford1997AJ}, \citet{Mei2012ApJ}). The thick shaded lines in the bottom of Figure \ref{fig:lynx-north} represent the non-zero cross shear component measured in each filter, where their thickness is the 1$\sigma$ width as measured in each bin and which are representative of the $\gamma_t$ (thin line) errors. The same distribution of galaxies is used in the measurements of tangential and cross shear. The broad agreement between the bands, which were independently obtained and analyzed, suggests that observational and modeling systematics which can vary between exposures are sub-dominant to our lensing signal.

The consistency of tangential shear between bands also implies that the weak lensing analysis can be improved by combining the information contained in multiple bands. Therefore, our shear analysis leverages the shape of each galaxy in three bands by providing an estimate of the error on shape measurement, wherein only galaxies with agreement between bands are used in later analysis including mass mapping and correlation function measurements. A traditional test of PSF and shape modeling error, the star-galaxy cross-correlation function, is described later subsection \ref{subsec:lynx-null_cross}.

\subsection{Mass mapping the Lynx field}\label{subsec:lynx_map}
Using the measurement of tangential and cross shear measured in each filter as input to Eqs. \ref{eq:s_t} and \ref{eq:s_x} we construct E- and B-mode aperture mass maps of the Lynx field. In the E-mode maps, the most massive structure in our field, the Lynx North cluster (previously seen in the maps of \citet{Miyazaki2007ApJ}) is detected at $S_t > 10$ in all three shape catalogs $V,R_c,i'$. The combined S/N map of Lynx North is shown in the top of Figure \ref{fig:lynx-north}, where contours colored from blue to red indicate $S_t$ from -3 to 8. The contours indicate the complexity of the mass clustering in the field of view and which can also be seen in the optical RGB image in that figure (composed from $i', R_c,V$ bands, respectively). Red sequence galaxies in this region, indicators of cluster membership, often underlie areas of positive $S_t$ across the entire field.

The computed $S_t$ mass maps in all three bands are quite similar, indicating agreement on the location and size of E-modes in the field. Calculating the Pearson correlation coefficient on the E-mode maps in the $V,R_c,i'$ bands gives $\rho_{R_c,i'} = 0.5156 \pm 0.0005 $, $\rho_{R_c,V} = .4789 \pm .0003$, and $\rho_{i',V} = .4859 \pm .0004$, where the errors have been estimated from bootstrap resampling. Additionally, our $S_\times$ (B-mode) maps in the $VRi'$ bands also show spatially correlated structure, with Pearson correlation coefficients between the B-mode maps in the $R_c/i',R_c/V$ and $i'/V$ bands calculated to be $\rho_{R_c,i'} = 0.289 \pm 0.004 $, $\rho_{R_c,V} = 0.363 \pm 0.003$, and $\rho_{i',V} = 0.221 \pm 0.004$, again with errors estimated from bootstrap resampling. Because the background shapes in multiple filters were measured independently using observations with large dithers and nights separated by many months as well as vastly different stellar PSF patterns, it is unlikely that PSF error is the origin of these cross-band B-modes.

\subsection{Comparison with HST observations}
\label{subsec:hst_map}
The Lynx field has been the subject of multiple previous investigations, including deep Hubble Space Telescope observations and a weak lensing analysis of ACS images presented in \citet{Jee2006ApJ}. In that paper, a shapelet decomposition method is used to measure the shear which aided in the $3\sigma$ detection of two $M\sim 2\times 10^{14}M_{\odot}$ members of the $z\sim 1.3$ super clustering in the field, Lynx-E and W, which have been well-studied and verified with Chandra X-ray analysis in addition to the weak lensing mass. The HST imagery we access unfortunately does not cover the central region of Lynx North at $z\sim0.55$, but its influence is easily observed in the shear field near the boundary. Indeed, the tangential shear about the coordinates of Lynx North (outside of the HST observations) is virtually identical with both the shears as measured in the Suprime-cam observations and the model shown in the bottom of Figure \ref{fig:lynx-north}. 

These space-based shapes provide a useful cross-check for our systematics in the ground-based observations. The stark differences in shape measurement method and PSF provide an outside test of the E- and B-mode detection in situations with varying image orientations, resolutions, cameras, optics, and atmospheric effects. We match this space-based shape catalog with the photometry provided by the Suprime-cam observations, and then map the E- and B-modes as was done in the ground-based Suprime-cam observations. These space and ground-based maps are then compared, and broad agreement is seen both visually and statistically: for the E-mode maps, the Pearson correlation coefficient between space and ground is $\rho_{E:space,ground}= 0.374 \pm 0.005$, and the B-mode maps are statistically correlated as well, $\rho_{B:space,ground}= 0.210 \pm 0.0004$. This space-based observation therefore validates the PSF modeling, shape measurement, and mass mapping algorithm used in the analysis of the ground-based data, indicating that the maps not biased by observational effects on the scales which are shared by both data. Additionally, such comparisons open up an exciting opportunity to utilize wide-field lensing maps (such as those from surveys) to model the external wide-field convergence of higher-redshift weak lensing observations on narrower fields of view (e.g. \citet{Fassnacht2006ApJ}). Indeed, use of the wide-field lensing map in the analysis of the deeper space-based catalog has provided a tentative detection of a filament between the two $z\sim 1.3$ clusters Lynx E and W.

\section{Lens modeling of the Lynx field}
\label{sec:simlynx}
\begin{figure}
  \includegraphics[width=\columnwidth]{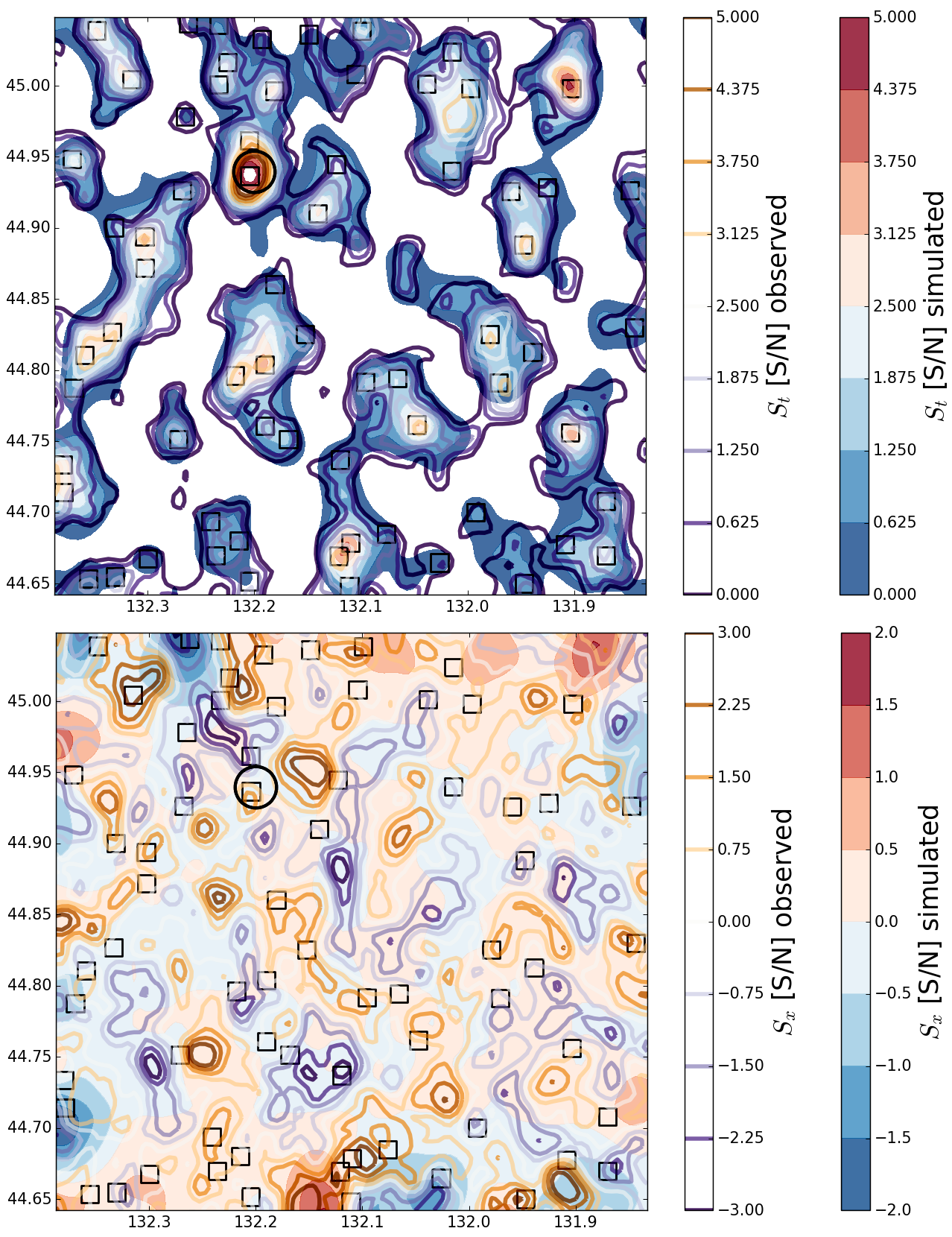}
  \caption{E-mode (top) and B-mode (bottom) maps comparing the simulated and observed maps as filled contours and lined contours respectively. Location of Lynx North is indicated by the circle, and other detected halos in the field above $S_t=2$ are shown as empty black squares. The detected peaks have their mass and redshift estimated, and their induced shear is summed on all background galaxies to produce the simulation estimate. Pearson correlation coefficients calculated using the simulated and observed maps indicate measurable correlation both E-modes ($\rho_{E:sim,obs}= 0.83 \pm 0.02$) and B-modes ($\rho_{B:sim,obs}= 0.22 \pm 0.01$), indicating generation of weak lensing B-modes by fields with multiple lenses.}
  \label{fig:lynx-ebmaps}
\end{figure}

To investigate the observed E- and B-modes in the Lynx field, we model the foreground lens field as we did in the $N$-body simulations of Section \ref{sec:simlens}. We do this by segmenting the E-mode aperture mass map and assigning clusters to locations to peaks in RA/Dec/z, and then simultaneously fitting $N_{clust}\sim100$ mass peaks as indicated by boxes in Figure \ref{fig:lynx-ebmaps}. As was previously shown, these peaks are correlated with the actual foreground lensing structure of clusters or groups of galaxies. Though some of these assumed peaks may not correspond to real halos, and the real halos may not have exactly NFW shear profiles, their strength, spatial distribution, and number density above $M=10^{13} M_{\odot}$ is similar to what is expected in similarly-sized fields (see Section \ref{sec:simlens}).

We then calculate the shear of each of the $\sim 10^4$ background galaxies with $z_{phot}>0.8$ as being the sum of shears induced by the estimated foreground halos. This is done for each background galaxy located in 3-dimensional coordinate space (RA/Dec/z) by minimizing the difference between the approximate and observed $S_t$ maps and computing the total reduced shear $g_{tot}$ as in Equation \ref{eq:shear_tot}. Since the simulations are shape-noise free, the approximate model has reduced shears which are mostly of an amplitude $g_{1,2}<0.1$ in correspondence with the weak lensing limit. Due to the lack of shape noise, the ellipticity components of the simulation are much smaller than the observations. However, the correlation is statistically significant, $\rho_{e1:sim,obs}= 0.084 \pm 0.004$ and $\rho_{e2:sim,obs}= 0.088 \pm 0.014$, where errors have been estimated from bootstrap resampling. Adding a shape noise (pre-lensing) broadens the $e_{1,2}$ distributions to the observed values, but unnecessarily degrades the correlation between simulated and observed shears and thus shape noise is not applied in these simulations.

These estimated shears can then be run with same aperture mass mapping algorithm as described in the previous section to produce estimated maps. The observed and approximated maps can be visually compared in Figure \ref{fig:lynx-ebmaps}. As in the $N$-body simulation analysis, the E-mode maps are very strongly similar as they have been iteratively fit to match the observed maps under the assumption that the majority of the E-modes are due to lensing, and the correlation coefficient of the simulated E-mode maps with the observed maps is $\rho_{E:sim,obs}= 0.83 \pm 0.02$. The B-mode maps exhibit weaker correlation by eye due to the dominance of shape noise over the lensing signal, but the correlation remains statistically significant with $\rho_{B:sim,obs}= 0.22 \pm 0.01$, even when edge effects are excluded.

\begin{figure}
  \includegraphics[width=\columnwidth]{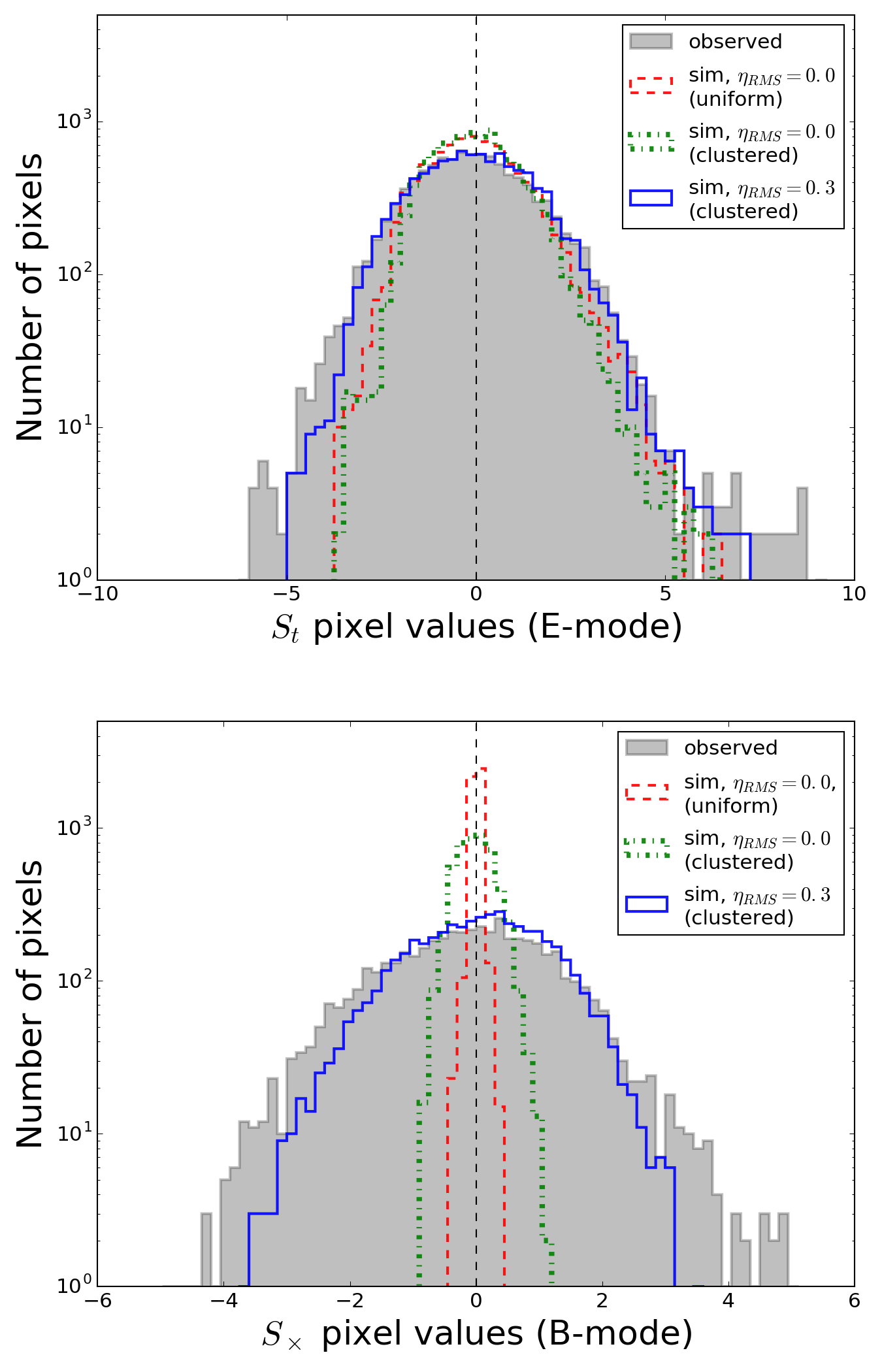}
  \caption{The distribution of E- and B-mode map pixel values. Red lines show the simulated shears of a uniform (homogeneous) source plane of galaxies at $z=1.0$ that fill the observed field of view. Green lines use the actual (inhomogeneous) distribution of source galaxies, which with the addition of shape noise form the blue lines, better match the observed distributions (shaded). B-mode distributions are shown without edge effects, indicating the presence of multiple lensing B-modes in the uniform galaxy simulation.}
  \label{fig:lynx-ebhist}
\end{figure}

As in Section \ref{sec:simlens}, we can use this model of foreground lenses to probe the effects of different source galaxy distributions. First, we can remove the source clustering B-modes by lensing a uniform background source galaxy plane (with the same number density of galaxies per square arcminute, but at a single redshift $z=1.0$). This is shown as the narrowest distribution of B-modes in Figure \ref{fig:lynx-ebhist} and indicates the sub-dominance of multiple lensing which is expected given the low density of background sources and lack of fortuitous extreme alignment between foreground clusters. Going beyond a uniform source plane by using the actual (measured) 3-dimensional distribution of background galaxies results in B-mode generation from the source clustering effect. This effect broadens the E- and B-mode distributions as it adds depth and variance to each. Finally, we can add shape noise to this clustered distribution of background galaxies and observe the broadening of the E- and B-mode pixel distributions shown as blue lines to nearly the widths of the observed pixels as shown by the filled histograms. This is the most realistic modeling of the observations, and includes all sources of known B-modes including double lensing, source clustering, and E-to-B mode mixing. The edge effects (which can be seen in the bottom panel of Figure \ref{fig:lynx-ebmaps}) are excluded from the histograms of Figure \ref{fig:lynx-ebhist} for clarity, though they too contain information about the locations and density of foreground lenses.

The flexibility of the lensing simulation is thus capable of encapsulating the known sources of B-modes from lensing and mass mapping, most of which can be attributed to shape noise, source lens clustering, and E-to-B mixing. However there are B-modes in the observed $V/R_c/i'$ maps which have larger amplitudes than can be accounted for by our lensing simulation. These high S/N B-modes, shown as the excess of gray shaded area over the blue line in Figure \ref{fig:lynx-ebhist}, could be due to uncorrected PSF effects. However, these high S/N peaks are consistent between bands with vastly different observing conditions. It therefore seems more likely that they might be accounted for with a more realistic simulation of the lensing potential distribution which we can only approximate. For instance, in the simulation we do not include void lensing, which would increase the number of negative E-modes and likely contribute positive and negative B-modes. However, void lensing was not accounted for in our approximation of $N$-body simulations and in that case the approximate B-modes distributions were an upper limit (see Figure \ref{fig:buzz-ebhist}). Further investigation of interesting high S/N peaks in B-mode maps is therefore warranted. Additionally, lensing on scales less than $\sim$ 1 Mpc and scales which span the entire field of view $\sim$ 100 Mpc are unaccounted for in our finite field reconstruction.  In fact, B-modes may be induced by large lenses outside the field which generate E-modes and induce wide-field correlations in background ellipticities.

\subsection{Shear correlation functions in the Lynx field}
\label{subsec:lynx_correl}
\begin{figure}
  \includegraphics[width=\columnwidth]{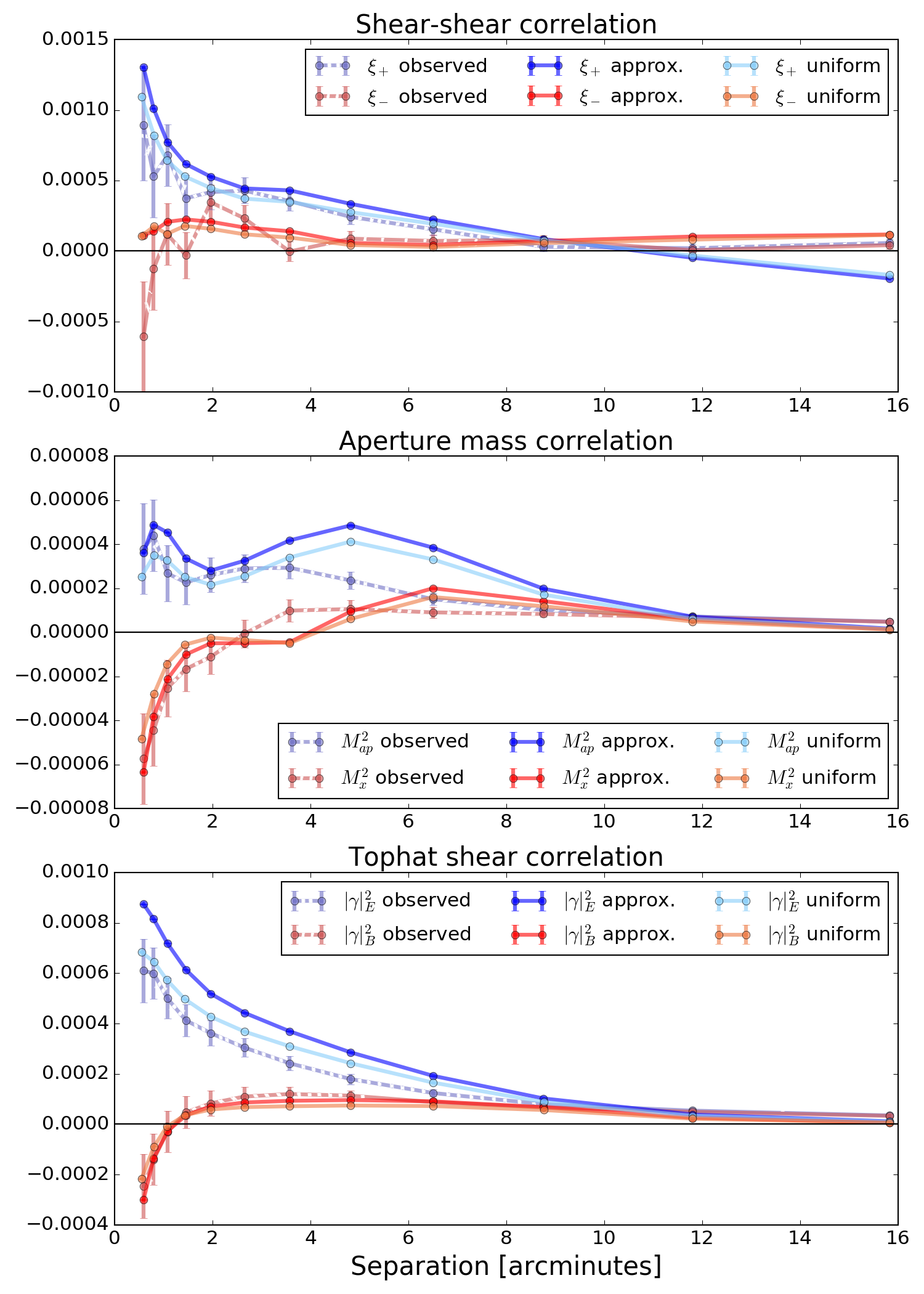}
  \caption{The three shear correlation functions for the Lynx field, $\xi_{\pm}$, $\langle M_{ap,\times}^2\rangle$, and $\langle \gamma_{E,B}^2\rangle$. In each panel, for each correlation function, three different sets of shears are correlated: 1) the observed shears and positions (which are clustered), 2) the simulated shears (which are lensed by multiple planes) and observed locations (in RA/Dec and z), and 3) a uniform background of galaxies at $z=1.0$ with the same number density as the observations. This shows that in the case of the correlation functions, the observed B-modes are not just due to the clustering.}
  \label{fig:lynx_correlations}
\end{figure}

Statistical correlations are again calculated to compliment our spatial maps. Correlation functions over the field are computed on scales of $0.5<\theta<15$ arcminutes using both the observed shears and those simulated under multiple conditions shown in Figure \ref{fig:lynx_correlations}. As in the $N$-body simulations, this selected field has an overabundance of large scale structure which give large amplitudes to the shear correlation functions as computed on this field. The shear correlations $\xi_{\pm}$, E- and B-mode aperture mass dispersion $\langle M_{ap,\times}^2 \rangle$, and tophat shear dispersion $\langle \gamma^2 \rangle_{E,B}$ are calculated under three different conditions: the observed shears, the shears which are simulated using the observed RA/Dec/z positions on the sky and estimated lens positions, and those simulated with uniform source plane at $z=1.0$ at the observed galaxy number density of $13 ~ \rm{arcmin}^{-2}$. E-modes for each statistic are the upper curves in each subplot in blue shades; B-modes are shown in shades of red. Error bars on the observed correlation functions are from pure shape noise with an $\eta_{rms}=0.3$, and the simulated cases have no shape noise added.

As in the $N$-body simulations of Section \ref{sec:simlens}, there are non-zero B-mode correlations on the scales shown, which ordinarily might be blamed on incomplete modeling of the PSF or gaps in the data. However, these B-modes are of similar amplitude in each band, and the simulated correlation function has no masking when a uniform lens plane is used. This implies that the majority of observed B-modes are not due to a PSF modeling issue or even the clustering of source galaxies as in the mapping case, but rather are intrinsic to these particular decompositions of a realistic shear field into E- and B-modes. On the smallest and largest scales, the finite depth and field of view our observations limit the size of E-modes which are measurable with our data, and this necessarily limits our E- and B-mode simulation which are constrained by the incompleteness of observations. The shear leakage from E- to B-modes on small scales, as discussed in Section \ref{subsec:buzzard-shear_corr}, is one such artifact of observation limitations, and though not astrophysically induced, still represent signal which is encapsulated in our shear model of the foreground lenses.

\subsection{Null tests and cross-validation}
\label{subsec:lynx-null_cross}
\begin{figure}
  \includegraphics[width=\columnwidth]{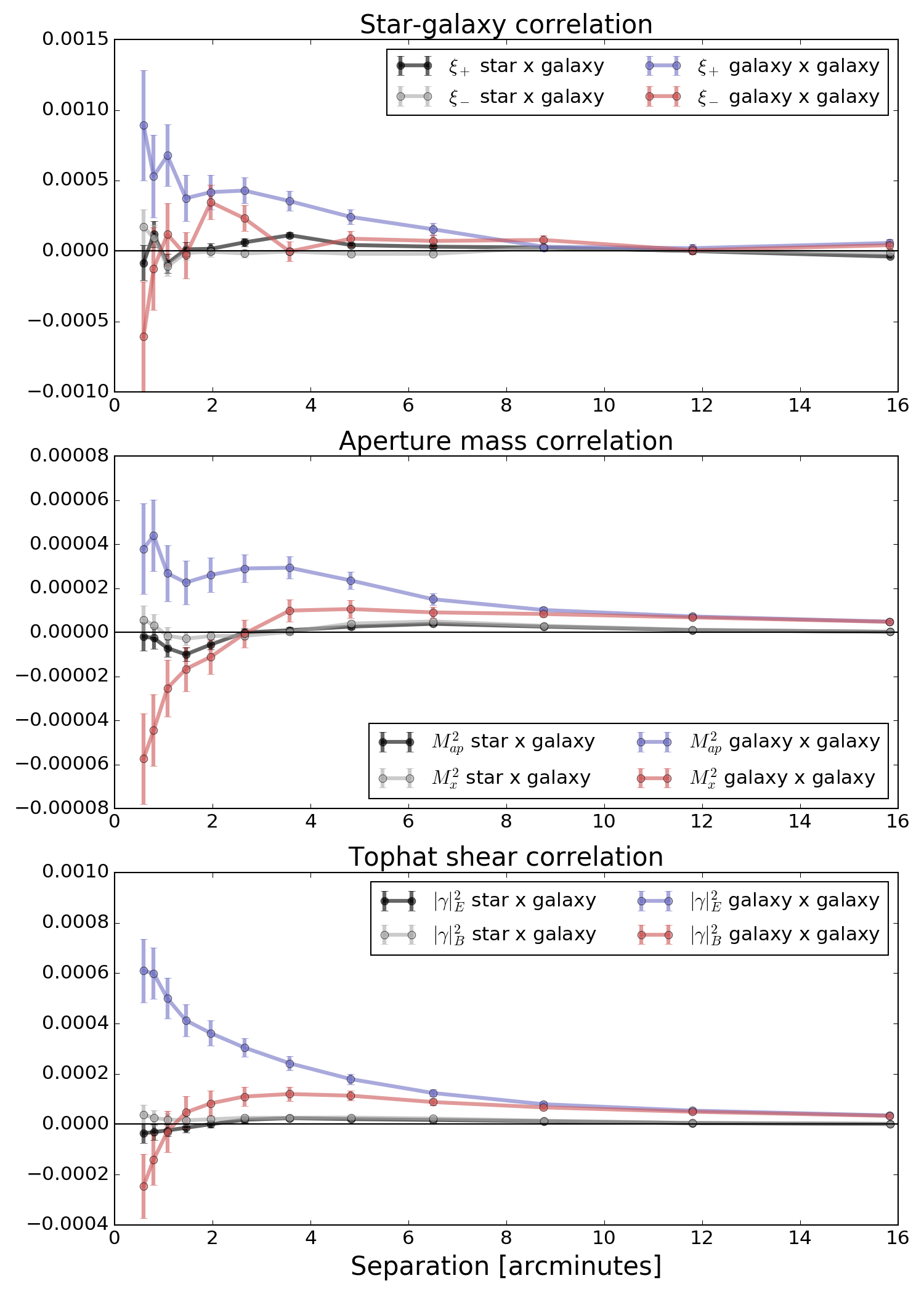}
  \caption{The star-galaxy cross correlation functions, $\xi_{\pm}$, $\langle M_{ap,\times}^2\rangle$, and $\langle \gamma_{E,B}^2\rangle$ are shown in black and gray, computed using the uncorrected ellipticities of PSF stars in the $R_c$ co-added image and the \texttt{sFIT} ellipticities of the background source galaxy sample used in the lensing analysis. For reference, the observed galaxy-galaxy shear correlation functions are shown in blue and red, as in Figure \ref{fig:lynx_correlations}, illustrating the minor contribution of PSF residuals to the observed E- and B-modes in our galaxy autocorrelation function analysis.}
  \label{fig:lynx_stargal_correlations}
\end{figure}

As a consistency check on PSF modeling and interpolation, we test for the presence of residual PSF systematics using the star-galaxy cross correlation function which is used in many weak lensing studies. These correlations can be computed using the same shear autocorrelation formalism presented in Section \ref{subsec:buzzard-shear_corr} but with the substitution of PSF ellipticity for one of the two shears. In this cross correlation, we use the uncorrected PSF ellipticity of stars $\epsilon_{uncorr}^\star$ which can reveal a correlated leakage of residuals from the PSF modeling into the shapes of galaxies. This test is presented in Figure \ref{fig:lynx_stargal_correlations}, where the grey and black lines show the ellipticities of stars in the co-added images in the $R_c$ filter that are cross-correlated with the ellipticities of the background ($z>0.8$) sample of galaxies used in the lensing analysis. In that figure we also show the observed galaxy shear auto-correlation as a reference, which greatly exceeds the signal in the star-galaxy cross correlation as measured in this way. 

\begin{figure}
  \includegraphics[width=\columnwidth]{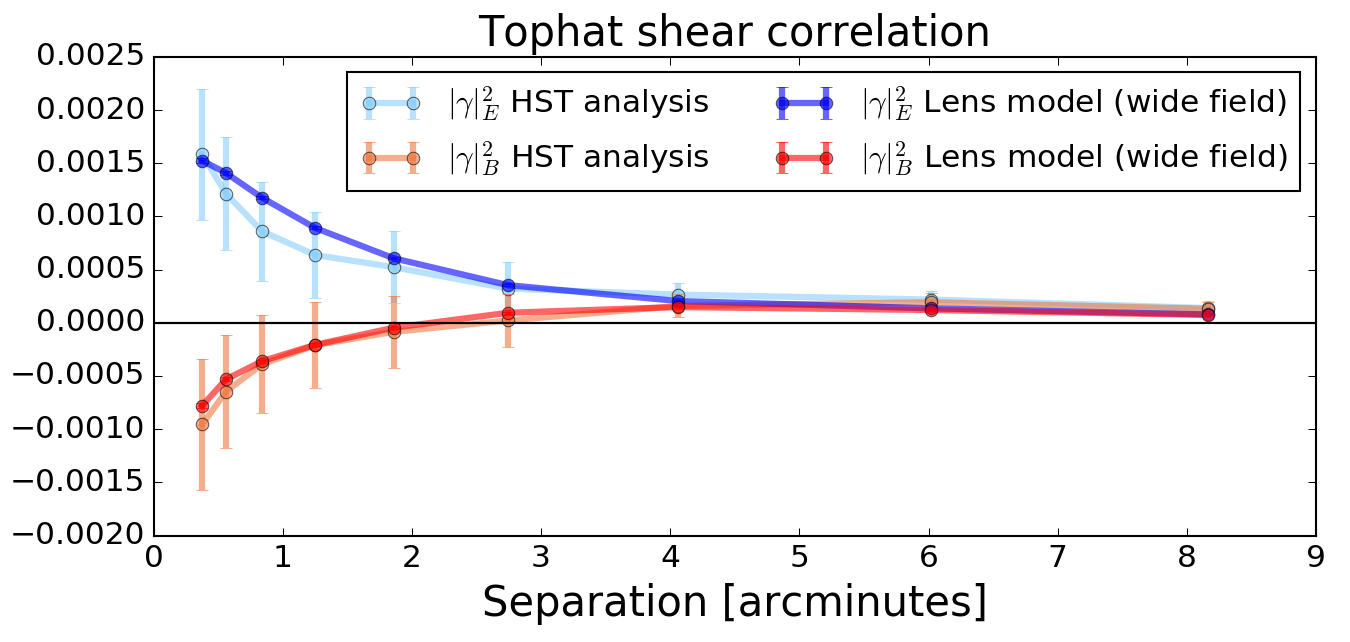}
  \caption{The top-hat shear correlation computed in the sub-field of the Suprime-cam observations which was observed with HST-ACS. The lens model shown uses the same halos as the wide field analysis, but computed at positions which were observed with HST. The lens model provides good agreement on the amplitude and scale of the E- and B-modes, even though the HST observations and analysis are significantly different from those of the ground-based Suprime-cam.}
  \label{fig:lynx_hst_vs_lens_model}
\end{figure}

Additionally, a type of cross-validation test can be performed using the overlapping HST-ACS observations (described in subsection \ref{subsec:hst_map}) which used a different camera \& optical system at a dissimilar orientation and resolution, without atmospheric effects, and processed using a different PSF and shape estimation routine \citep{Jee2006ApJ}. In Figure \ref{fig:lynx_hst_vs_lens_model}, we compare the galaxy shapes measured using the HST-ACS observations to our wide-field lens model evaluated at those galaxy positions using the top-hat filtered shear correlation. As can be seen in that figure, both the lens model and HST-ACS filtered correlation functions are in agreement about the scale and amplitude of the E- and B-modes in this subset of the data. Interestingly, though the field of view of the HST-ACS mosaic is narrow, it probes a region of high shear near the Lynx-North cluster. This high-shear region shows an increased E-mode, and a complementary increase in the B-mode, in both the lens model and the HST observations.

These cross-correlation and cross-validation tests, in addition to the aforementioned consistency of E- and B-modes between the independent $V,R,i'$ filters, implies (but does not prove) that our lensing analysis is not dominated by PSF misestimation or other observational systematic errors. 

\section{Discussion}
\label{sec:discussion}
Our shear modeling method has been applied in the analysis of the Buzzard $N$-body simulations and a field of galaxy clusters in Lynx. We described the method of mass mapping in Section \ref{sec:massmap_method} using the shears from ray-tracing the Buzzard $N$-body simulations. Section \ref{sec:simlens} introduced the method of approximating the ray-traced shear field as a sum of finite foreground cluster lenses, which is tested by comparison of the ray-traced shears to those computed using the known halo catalog and assumed NFW shear profile. Using this approximation, we explain the observation of B-modes in the spatial maps and shear correlations in the $N$-body simulation sub-field. These apparent B-modes can be further understood using the flexibility of our method which allows for modular variation of the spatial distribution of lenses and galaxies. For instance, changing the positions of lensed galaxies from a clustered distribution to a uniform one at fixed redshift reduces much of the B-mode signature. Similarly, collapsing all lenses to a single redshift also eliminates the B-mode signature of double (or multiple) lensing in cases with higher source densities of galaxies. Because these $N$-body simulation B-modes cannot be systematic errors resulting from  shape noise, PSF mis-estimation, or other observational systematics, they must either result from the lensing field or the nature of shear analysis. In either case, the generated B-modes represent a measurable signal when galaxies which are realistically distributed amongst, and then lensed by, an inhomogeneous 3-dimensional web of lenses.

Using this flexible approximation to the lensing potential we can begin to account for the known sources of lensing \& mapping B-modes, using them as a signal unto themselves and as an opportunity for discovery. In Section \ref{sec:data} we apply our approximate lens modeling technique to deep observations of a field of galaxy clusters with the goal of modeling the observed E- and B-mode maps and correlations. We described our data reduction process, which goes from raw observational data to photometric redshift calibration and shear measurement using the `stack-fit' algorithm on multiple dithered exposures and multiple bands. After aperture mass mapping and correlating these shears, similar E- and B-modes were observed in all three filters as well as in overlapping deep HST observations. These space-based measurements provide an independent confirmation of the observed B-modes using a different shape measurement algorithms, camera \& optics, and without atmospheric effects. As in the $N$-body simulations, most of the non-shape noise B-modes can be attributed to source lens clustering and E-to-B mixing, both of which depend on the distribution and density of foreground lenses. This approximate method can also be used to de-lens the effect of wide-field foreground lenses on higher redshift clusters, effectively accounting for external convergence in deeper observations on narrower fields of view.

Interestingly, the deep observations also contain B-modes in common to the $V/R_c/i'$ maps which have larger (signal to noise) amplitudes than can be accounted for by our lensing approximation. It therefore seems likely that these modes might be accounted for with a more realistic simulation of the lensing potential. For instance, inclusion of the lensing by voids (underdensities) along the line of sight would increase the number of negative E-modes and likely contribute positive and negative B-modes. Additionally, lensing on scales smaller and larger than our observations allow, those scales less than $\sim$ 1 Mpc and greater than $\sim$ 100 Mpc, are also unaccounted for in our finite field reconstruction. B-modes could therefore also be observationally induced by large lenses outside the field which generate E-modes and induce wide-field intrinsic shears in source galaxies. Therefore, the variance in our estimated B-mode distributions must be taken as lower limits. However, lensing by voids and larger-scale structure was not included in our approximation of $N$-body ray-traced shears and yet all high S/N peaks were accounted for in those simulations. This discrepancy between $N$-body simulations and observations warrants further investigation, and follow-up studies are underway using the wider fields available in modern weak lensing surveys and other $N$-body simulations. 

\section{Summary}
\label{sec:summary}
In this paper we have explored mass mapping and shear correlations using measurements of shears and redshifts from $N$-body simulations and multi-band observations similar to what will be available with the LSST 10 year dataset. Data is processed similarly in both observations and simulations, concluding with aperture mass mapping and shear correlation function measurement. We developed and tested a model for representing the shear as the successive lensings in a piece of the cosmic web by associating E-mode overdensities with galaxy clusters along the line of sight. This simple approximation to $N$-body ray tracing is surprisingly useful at capturing the shear field in the weak lensing limit, even when the mass and 3-d positions of clusters can only be estimated from observational data. We compare this model to observations through both aperture mass maps and shear correlations and demonstrate that the decomposition of the observed shear field into gradient (E-modes) and curl (B-modes) yields general agreement with the model, demonstrating how the patterns of pure tangential shear produced by realistic distributions of lenses \& galaxies also induce a measurable B-mode signal. Contrary to a systematic error, these observational B-modes contain information about the lenses in the field. Further application of this method of approximation on real data therefore presents an opportunity to use B-modes as an aid for discovery and a signal unto itself, especially using wide field weak lensing surveys such as CFHTLenS, DES, KiDS, Euclid, and the LSST.

\section*{Acknowledgements}
We thank Michael Schneider and Sam Schmidt for many helpful discussions, as well as the anonymous referee for valuable feedback which helped clarify the text. Financial support from DOE grant DE-SC0009999 and Heising-Simons Foundation grant 2015-106 are gratefully acknowledged. We thank Risa Wechsler, Joe DeRose, and the Buzzard simulation team for their N-body lensing simulation catalogs. M.J.J. acknowledges support for the current research from the National Research Foundation of Korea under the programs 2017R1A2B2004644 and 2017R1A4A1015178.


\bibliographystyle{mnras}
\bibliography{lynx} 


\bsp	
\label{lastpage}
\end{document}